\documentclass[twocolumn,aip]{revtex4}
\usepackage{natbib}
\usepackage{amsmath}
\usepackage{amssymb}
\usepackage{graphicx}
\usepackage{amsbsy}
\usepackage{bm}

\newcommand{\refeq}[1]{(\ref{#1})}

\newcommand{\bA}{{\bf A}}

\newcommand{\bB}{{\bf B}}

\newcommand{\bD}{{\bf D}}

\newcommand{\be}{{\bf e}}
\newcommand{\bF}{{\bf F}}
\newcommand{\bff}{{\bf f}}

\newcommand{\bI}{{\bf I}}
\newcommand{\bj}{{\bf j}}

\newcommand{\bk}{{\bf k}}

\newcommand{\bL}{{\bf L}}
\newcommand{\bn}{{\bf n}}

\newcommand{\bQ}{{\bf Q}}
\newcommand{\br}{{\bf r}}
\newcommand{\bR}{{\bf R}}

\newcommand{\bT}{{\bf T}}

\newcommand{\bv}{{\bf v}}
\newcommand{\bV}{{\bf V}}
\newcommand{\bu}{{\bf u}}
\newcommand{\bU}{{\bf U}}
\newcommand{\bw}{{\bf w}}
\newcommand{\bW}{{\bf W}}

\newcommand{\bX}{{\bf X}}

\newcommand{\bz}{{\bf z}}

\newcommand{\bOmega}{\mbox{\boldmath$\Omega$}}

\newcommand{\brho}{\boldsymbol{\rho}}

\newcommand{\bnabla}{\mbox{\boldmath$\nabla$}}
\newcommand{\btimes}{\boldsymbol{\times}}
\newcommand{\bcdot}{\,\mbox{\boldmath{$\cdot$}}\,}

\newcommand{\eff}{{\rm eff}}

\newcommand{\llarge}{{\rm L}}

\newcommand{\ssmall}{{\rm S}}

\newcommand{\diff}{{\,\mathrm d}}

\newcommand{\im}{{\mathrm i}}

\newcommand{\smallfrac}[2]{{\textstyle\frac{#1}{#2}}}
\newcommand{\halff}{{\textstyle\frac{1}{2}}}
\newcommand{\third}{{\textstyle\frac{1}{3}}}

\newlength{\tightsecu}
\newlength{\tightsecl}
\newlength{\tightsubsecu}
\newlength{\tightsubsecl}
\newlength{\tightsubsubsecu}
\newlength{\tightsubsubsecl}
\newlength{\sectosubsec}
\newlength{\subsectosubsubsec}
\newcommand{\mathsfb}{\mathsf}
\newcommand{\mathsfbi}{\mathsf}

\newcommand{\remark}[1]{{\tt[#1]}}

\newcommand{\fluid}{{\textrm{f}}}
\newcommand{\suspension}{{\textrm{s}}}
\newcommand{\pressure}{{\textrm{p}}}
\newcommand{\force}{{\textrm{f}}}
\newcommand{\particle}{{\textrm{p}}}
\newcommand{\periodic}{{\textrm{per}}}
\newcommand{\asymptotic}{{\textrm{HS}}}
\newcommand{\dipole}{{\textrm{d}}}
\newcommand{\surface}{{\textrm{s}}}
\newcommand{\lowDensity}{0}
\newcommand{\effective}{{\textrm{eff}}}
\newcommand{\noSlip}{{\textrm{NS}}}
\newcommand{\pointForce}{0}
\newcommand{\pointTorque}{0}
\newcommand{\oneParticle}{1}

\newcommand{\unitCell}{{\Omega_{\textrm{p}}}}
\newcommand{\unitCellArea}{{S_{\textrm{p}}}}
\newcommand{\wholeSpace}{{\Omega_\infty}}
\newcommand{\unitTensor}{{\mathsfb{I}}}
\newcommand{\lateralUnitTensor}{\,\unitTensor_\parallel}
\newcommand{\lateralAlternatingTensor}{{\boldsymbol{\epsilon}}_\parallel}

\newcommand{\appliedForce}{\totForce}
\newcommand{\appliedTorque}{\totTorque}
\newcommand{\appliedPressureGradient}{\bnablaLat\externalPressure}

\newcommand{\particleVelocity}{\bU}
\newcommand{\particleAngularVelocity}{{\boldsymbol{\Omega}}}

\newcommand{\generalTensorCoefficient}{\psi}
\newcommand{\generalTensor}{\boldsymbol{\generalTensorCoefficient}}

\newcommand{\resistanceCoefficient}{\zeta}
\newcommand{\resistanceMatrix}{{\boldsymbol{\resistanceCoefficient}}}
\newcommand{\resistanceMatrixTP}{\resistanceMatrix^{\transl\parabolic}}
\newcommand{\resistanceMatrixRP}{\resistanceMatrix^{\rot\parabolic}}
\newcommand{\resistanceMatrixPT}{\resistanceMatrix^{\parabolic\transl}}
\newcommand{\resistanceMatrixPR}{\resistanceMatrix^{\parabolic\rot}}
\newcommand{\resistanceMatrixPP}{\resistanceMatrix^{\parabolic\parabolic}}
\newcommand{\resistanceCoefficientPP}
    {\resistanceCoefficient^{\parabolic\parabolic}}
\newcommand{\normalizedResistanceCoefficientPP}
    {\tilde{\resistanceCoefficient}^{\parabolic\parabolic}_\oneParticle}
\newcommand{\averageResistanceCoefficientPP}
           {{\bar\resistanceCoefficient}^{\parabolic\parabolic}}

\newcommand{\averageResistanceMatrixPP}
           {{\bar{\resistanceMatrix}}^{\parabolic\parabolic}}

\newcommand{\mobilityCoefficient}{\mu}
\newcommand{\mobilityMatrix}{{\boldsymbol{\mobilityCoefficient}}}
\newcommand{\mobilityMatrixP}{\mobilityMatrix}

\newcommand{\mobilityCoefficientPP}
           {\mobilityCoefficient^{\parabolic\parabolic}}
\newcommand{\mobilityCoefficientPT}
           {\mobilityCoefficient^{\parabolic\transl}}
\newcommand{\mobilityCoefficientPR}
           {\mobilityCoefficient^{\parabolic\rot}}

\newcommand{\normalizedMobilityCoefficientPP}
    {\tilde{\mobilityCoefficient}^{\parabolic\parabolic}_\oneParticle}

\newcommand{\mobilityMatrixPP}{\mobilityMatrixP^{\parabolic\parabolic}}
\newcommand{\mobilityMatrixTP}{\mobilityMatrixP^{\transl\parabolic}}
\newcommand{\mobilityMatrixRP}{\mobilityMatrixP^{\rot\parabolic}}

\newcommand{\mobilityMatrixPT}{\mobilityMatrixP^{\parabolic\transl}}
\newcommand{\mobilityMatrixPR}{\mobilityMatrixP^{\parabolic\rot}}

\newcommand{\mobilityMatrixTT}{\mobilityMatrix^{\transl\transl}}
\newcommand{\mobilityMatrixTR}{\mobilityMatrix^{\transl\rot}}
\newcommand{\mobilityMatrixRT}{\mobilityMatrix^{\rot\transl}}
\newcommand{\mobilityMatrixRR}{\mobilityMatrix^{\rot\rot}}

\newcommand{\averageMobilityMatrix}{{\bar\mobilityMatrix}}
\newcommand{\averageMobilityMatrixPP}
           {\averageMobilityMatrix^{\parabolic\parabolic}}
\newcommand{\averageMobilityMatrixPT}
           {\averageMobilityMatrix^{\parabolic\transl}}
\newcommand{\averageMobilityMatrixPR}
           {\averageMobilityMatrix^{\parabolic\rot}}
\newcommand{\averageMobilityMatrixTP}
           {\averageMobilityMatrix^{\transl\parabolic}}
\newcommand{\averageMobilityMatrixTT}
           {\averageMobilityMatrix^{\transl\transl}}
\newcommand{\averageMobilityMatrixTR}
           {\averageMobilityMatrix^{\transl\rot}}
\newcommand{\averageMobilityMatrixRP}
           {\averageMobilityMatrix^{\rot\parabolic}}
\newcommand{\averageMobilityMatrixRT}
           {\averageMobilityMatrix^{\rot\transl}}
\newcommand{\averageMobilityMatrixRR}
           {\averageMobilityMatrix^{\rot\rot}}

\newcommand{\pointForceEffectiveMobilityCoefficient}
          {\effectiveMobilityCoefficient_\pointForce^{\parabolic\transl}}
\newcommand{\pointTorqueEffectiveMobilityCoefficient}
          {\effectiveMobilityCoefficient_\pointTorque^{\parabolic\rot}}

\newcommand{\effectiveMobilityCoefficient}{\nu}
\newcommand{\effectiveMobilityMatrix}
   {{\boldsymbol{\effectiveMobilityCoefficient}}}

\newcommand{\effectiveMobilityMatrixTP}
  {\effectiveMobilityMatrix^{\transl\parabolic}}
\newcommand{\effectiveMobilityMatrixTT}
  {\effectiveMobilityMatrix^{\transl\transl}}
\newcommand{\effectiveMobilityMatrixTR}
  {\effectiveMobilityMatrix^{\transl\rot}}

\newcommand{\effectiveMobilityMatrixRP}
  {\effectiveMobilityMatrix^{\rot\parabolic}}
\newcommand{\effectiveMobilityMatrixRT}
  {\effectiveMobilityMatrix^{\rot\transl}}
\newcommand{\effectiveMobilityMatrixRR}
  {\effectiveMobilityMatrix^{\rot\rot}}

\newcommand{\effectiveMobilityMatrixPP}
  {\effectiveMobilityMatrix^{\parabolic\parabolic}}
\newcommand{\effectiveMobilityMatrixPT}
  {\effectiveMobilityMatrix^{\parabolic\transl}}
\newcommand{\effectiveMobilityMatrixPR}
  {\effectiveMobilityMatrix^{\parabolic\rot}}

\newcommand{\effectiveMobilityCoefficientPP}
  {\effectiveMobilityCoefficient^{\parabolic\parabolic}}
\newcommand{\effectiveMobilityCoefficientPT}
  {\effectiveMobilityCoefficient^{\parabolic\transl}}
\newcommand{\effectiveMobilityCoefficientPR}
  {\effectiveMobilityCoefficient^{\parabolic\rot}}

\newcommand{\averageSuspensionVelocity}{{\bar\bu}}
\newcommand{\averageParticleVelocity}{{\bar\bU}}
\newcommand{\averageParticleAngularVelocity}{{\bar\particleAngularVelocity}}

\newcommand{\averagePointForceFlow}{{\bar\bu}_0}
\newcommand{\pointForceFlow}{\bv_0}

\newcommand{\suspensionDensityA}{{\bar n}_\surface}
\newcommand{\areaFraction}{\phi_\surface}
\newcommand{\suspensionDensityV}{n}

\newcommand{\averagePressure}{\bar p}
\newcommand{\macroscopicPressureGradient}{\bnablaLat\averagePressure}

\newcommand{\permeability}{{\boldsymbol{\kappa}}}
\newcommand{\scalarPermeability}{\kappa}
\newcommand{\channelPermeability}{\kappa_0}

\newcommand{\effectiveWidthPermeability}{\scalarPermeability_\effective}
\newcommand{\Heffective}{H_\effective}
\newcommand{\fluidFluxRatio}{\lambda}

\newcommand{\zNoSlip}{Z_\noSlip}

\newcommand{\periodicGreenT}{\bT^\periodic}
\newcommand{\GreenT}{\bT}
\newcommand{\GreenComponentT}{T}
\newcommand{\OseenTensor}{\GreenT_0}
\newcommand{\wallGreenT}{\GreenT'}
\newcommand{\wallGreenComponentT}{\GreenComponentT'}
\newcommand{\GreenTasymptotic}{\bT_\asymptotic}
\newcommand{\periodicGreenTasymptotic}{\bT_\asymptotic^\periodic}
\newcommand{\periodicGreenQ}{\bQ^\periodic}
\newcommand{\GreenQ}{\bQ}
\newcommand{\GreenComponentQ}{Q}
\newcommand{\OseenQ}{\GreenQ_0}
\newcommand{\wallGreenQ}{\GreenQ'}
\newcommand{\wallGreenComponentQ}{\GreenComponentQ'}
\newcommand{\GreenQasymptotic}{\bQ_\asymptotic}
\newcommand{\periodicGreenQasymptotic}{\bQ_\asymptotic^\periodic}

\newcommand{\Coulomb}{\ScalarBasisM{0}}
\newcommand{\Wigner}{w}
\newcommand{\dipolarPressure}{p^\dipole}
\newcommand{\dipolarFlow}{\bv^\dipole}

\newcommand{\dipoleMoment}{\bD}
\newcommand{\particleDipoleMoment}{\bD}
\newcommand{\averageParticleDipoleMoment}{\bar\bD}
\newcommand{\pressureDipoleMoment}{\bD_0}
\newcommand{\lateralPointForce}{\bF_\parallel}

\newcommand{\pointForceDipoleMomentCoefficient}
          {\mobilityCoefficient_\pointForce^{\parabolic\transl}}
\newcommand{\pointTorqueDipoleMomentCoefficient}
          {\mobilityCoefficient_\pointTorque^{\parabolic\rot}}

\newcommand{\STmatrix}{\mathsfb{B}}
\newcommand{\Qmatrix}{\mathsfb{b}}

\newcommand{\wall}{{\rm w}}
\newcommand{\Wall}{{\rm W}}
\newcommand{\up}{{\rm U}}
\newcommand{\low}{{\rm L}}
\newcommand{\incom}{{\rm in}}
\newcommand{\outcom}{{\rm out}}
\newcommand{\TW}{{\rm TW}}
\newcommand{\transl}{{\rm t}}
\newcommand{\rot}{{\rm r}}
\newcommand{\parabolic}{\rm p}

\newcommand{\as}{{\rm as}}
\newcommand{\cyl}{{\rm cyl}}

\newcommand{\ex}{\hat\be_x}
\newcommand{\ey}{\hat\be_y}
\newcommand{\ez}{\hat\be_z}
\newcommand{\eSpherical}{{\hat\be}}

\newcommand{\lateralDistance}{\rho}
\newcommand{\lateralVector}{\brho}

\newcommand{\identityTensor}{\hat\bI}

\newcommand{\bnablaLat}{\bnabla_\parallel}
\newcommand{\nablaLat}{\nabla_{\parallel}}

\newcommand{\e}{{\rm e}}
\newcommand{\BesselJ}{{\rm J}}

\newcommand{\totForce}{{\boldsymbol{\mathcal{F}}}}
\newcommand{\totTorque}{{\boldsymbol{\mathcal{T}}}}

\newcommand{\externalVelocity}{\bv^{\rm ext}}
\newcommand{\scatteredVelocityPart}{\bv'}

\newcommand{\externalPressure}{p^{\rm ext}}
\newcommand{\scatteredPressurePart}{p'}

\newcommand{\sphericalBasisP}[1]{\bv^+_{#1}}
\newcommand{\sphericalBasisM}[1]{\bv^-_{#1}}

\newcommand{\reciprocalSphericalBasisP}[1]{\bw^+_{#1}}

\newcommand{\ScalarBasis}[1]{\Phi_{#1}}
\newcommand{\periodicScalarBasis}[1]{{\tilde\Phi}_{#1}}
\newcommand{\ScalarBasisP}[1]{\ScalarBasis{#1}^+}
\newcommand{\ScalarBasisM}[1]{\ScalarBasis{#1}^-}
\newcommand{\periodicScalarBasisM}[1]{\periodicScalarBasis{#1}^-}

\newcommand{\ScalarBasisPcon}[1]{\ScalarBasis{#1}^{+\,*}}

\newcommand{\HeleShawBasis}[2]{\bv^{\as\,#1}_{#2}}

\newcommand{\GreenWallTotElement}{G}

\newcommand{\GrandFrictionElement}{F}

\newcommand{\ZsingleWall}{{\mathsfb Z}_\wall}

\newcommand{\tildeCartesianDisplacement}[1]{{\tilde{\mathsfb S}}_{\rm C}^{#1}}

\newcommand{\periodicScalarDisplacementElementsPM}{{\tilde S}_\cyl^{+-}}

\newcommand{\TransformationElementSAs}{C}

\newcommand{\matrixElementBPM}[1]{B^{\pm}_{#1}}

\newcommand{\tildeZW}{{\tilde{\mathsfbi Z}}_\TW}

\newcommand{\resistanceMatrixElement}{{\zeta}}
\newcommand{\resistanceMatrixTT}{\resistanceMatrix^{\transl\transl}}
\newcommand{\resistanceMatrixTR}{\resistanceMatrix^{\transl\rot}}
\newcommand{\resistanceMatrixRT}{\resistanceMatrix^{\rot\transl}}
\newcommand{\resistanceMatrixRR}{\resistanceMatrix^{\rot\rot}}

\newcommand{\twoParticleResistanceNorm}{\bar\resistanceMatrixElement}

\newcommand{\frictionProjectionVector}{\bX}

\begin{document}
\date{\today}

\title{An analysis of the far-field response to external forcing of a
suspension in Stokes flow in a parallel-wall channel}
\author{J.\ B{\l}awzdziewicz}
\affiliation{Department of Mechanical Engineering, Yale University,
P.O. Box 20-8286, New Haven, CT 06520}
\author{E.\ Wajnryb}
\affiliation{IPPT, \'Swi\c{e}tokrzyska 21, Warsaw, Poland}  

\begin{abstract}

The leading-order far-field scattered flow produced by a particle in a
parallel-wall channel under creeping flow conditions has a form of the
parabolic velocity field driven by a 2D dipolar pressure distribution.
We show that in a system of hydrodynamically interacting particles,
the pressure dipoles contribute to the macroscopic suspension flow in
a similar way as the induced electric dipoles contribute to the
electrostatic displacement field.  Using this result we derive
macroscopic equations governing suspension transport under the action
of a lateral force, a lateral torque or a macroscopic pressure
gradient in the channel.  The matrix of linear transport coefficients
in the constitutive relations linking the external forcing to the
particle and fluid fluxes satisfies the Onsager reciprocal relation.
The transport coefficients are evaluated for square and hexagonal
periodic arrays of fixed and freely suspended particles, and a simple
approximation in a Clausius-Mossotti form is proposed for the channel
permeability coefficient.  We also find explicit expressions for
evaluating the periodic Green's functions for Stokes flow between two
parallel walls.

\end{abstract}

\maketitle

\section{Introduction}
\label{Introduction}

There have been numerous studies of the effect of confinement on the dynamics
of rigid particles
\cite{
Staben-Zinchenko-Davis:2003,%
Jones:2004,%
Bhattacharya-Blawzdziewicz-Wajnryb:2005,%
Bhattacharya-Blawzdziewicz-Wajnryb:2005a,%
Bhattacharya-Blawzdziewicz-Wajnryb:2006a,%
Staben-Zinchenko-Davis:2006,%
Han-Alsayed-Nobili-Zhang-Lubensky-Yodh-2006,%
Zurita_Gotor-Blawzdziewicz-Wajnryb:2007a%
}, 
deformable drops
\cite{
Pathak-Migler:2003,%
Sibillo-Pasquariello-Simeone-Cristini-Guido:2006,%
Griggs-Zinchenko-Davis:2007,%
Janssen-Anderson:2007%
},
and macromolecules
\cite{%
Chen-Graham-de_Pablo-Randall-Gupta-Doyle:2004,%
Usta-Butler-Ladd:2007%
}
in creeping flow in a parallel-wall channel. The investigations revealed new
confinement-induced phenomena such as migration of macromolecules away from
the walls
\cite{
Chen-Graham-de_Pablo-Randall-Gupta-Doyle:2004,%
Jendrejack-Schwartz-de_Pablo-Graham:2004,%
Khare-Graham-de_Pablo:2006,%
Hernandez_Ortiz-de_Pablo-Graham:2006%
},
stability of strongly elongated drops in a confined shear flow
\cite{Pathak-Migler:2003,%
Sibillo-Pasquariello-Simeone-Cristini-Guido:2006,%
Janssen-Anderson:2007},
and cross-streamline migration of spherical particles due to pair
encounters in wall presence
\cite{%
Zurita_Gotor-Blawzdziewicz-Wajnryb:2007b%
}.
Confinement-related collective phenomena include spontaneous formation of
string-like drop configurations 
\cite{Pathak-Migler:2003,%
Sibillo-Pasquariello-Simeone-Cristini-Guido:2006%
},
propagation of displacement waves in linear trains of drops in Poiseuille flow
\cite{Beatus-Tlusty-Bar_Ziv:2006,Baron-Blawzdziewicz-Wajnryb:2008},
instabilities of confined particle jets \cite{Alvarez-Clement-Soto:2006}, and
pattern formation and rearrangements of particle lattice in 2D
regular particle arrays \cite{Baron-Blawzdziewicz-Wajnryb:2008}.  However,
while the above studies revealed rich dynamics resulting from hydrodynamic
interactions of particles with the channel walls, a comprehensive
understanding of mechanisms underlying the confinement effects is still
lacking.

Confinement-induced multiparticle collective phenomena often emerge as
a result of hydrodynamic interactions associated with the far-field
form of the flow produced by the particles moving in the channel
\cite{%
Cui-Diamant-Lin-Rice:2004,
Beatus-Tlusty-Bar_Ziv:2006,
Alvarez-Clement-Soto:2006,
Bhattacharya-Blawzdziewicz-Wajnryb:2006,
Baron-Blawzdziewicz-Wajnryb:2008}.  
Due to confinement, this flow qualitatively differs from the far-field
flow caused by particle motion in free space.  The difference stems
from the strong fluid-volume conservation constraint associated with
the wall presence, and from absorption of momentum by the walls.
Owing to the momentum absorption, the velocity field decays too fast
to produce a nonzero fluid flux through the boundary at infinity.
Therefore, the fluid displaced by a moving particle creates a backflow
pattern 
\cite{Liron-Mochon:1976,
Cui-Diamant-Lin-Rice:2004,
Bhattacharya-Blawzdziewicz-Wajnryb:2005a,
Bhattacharya-Blawzdziewicz-Wajnryb:2006} 
in order to ensure fluid incompressibility.  In contrast, in unbounded
systems the fluid in the whole space moves in the same direction as
the particle.

The far-field backflow produced by a particle in a channel has a form
of a parabolic Hele--Shaw flow driven by a 2D dipolar pressure
distribution
\cite{Liron-Mochon:1976,
Cui-Diamant-Lin-Rice:2004,
Bhattacharya-Blawzdziewicz-Wajnryb:2005a,
Bhattacharya-Blawzdziewicz-Wajnryb:2006%
}.   
The Hele--Shaw form of the flow far from the particle can be derived
using an appropriate lubrication expansion
\cite{Bhattacharya-Blawzdziewicz:2008}.  The dipolar character of the
velocity field around a spherical particle results from the
cylindrical symmetry of the problem.

An immediate consequence of the backflow effect is the negative sign
of the transverse component of the two-particle mutual
hydrodynamic-mobility coefficient
\cite{%
Cui-Diamant-Lin-Rice:2004,%
Bhattacharya-Blawzdziewicz-Wajnryb:2005%
}.
The backflow also causes a large resistance of elongated particles in
a narrow channel
\cite{%
Bhattacharya-Blawzdziewicz-Wajnryb:2005,
Bhattacharya-Blawzdziewicz-Wajnryb:2005a,
Bhattacharya-Blawzdziewicz-Wajnryb:2006,
Han-Alsayed-Nobili-Zhang-Lubensky-Yodh-2006%
}.
Furthermore, collective action of the dipolar flow fields produced by
individual particles gives rise to propagation of
particle-displacement waves in linear particle arrays in Poiseuille
flow \cite{Beatus-Tlusty-Bar_Ziv:2006}, and it governs macroscopic
deformation and lattice rearrangements in regular particle arrays
\cite{Baron-Blawzdziewicz-Wajnryb:2008}.

The dipolar far-field flow produced by the particles in a channel is
crucial not only for understanding suspension dynamics on the particle
scale, but also for describing the macroscopic fluid and particle
transport.  We show that particle contribution to the macroscopic
volume flux in suspension flow through the channel can be determined
from the amplitudes of the dipolar Hele--Shaw far-field flows produced
by the particles.  This behavior has a close physical analogy in
electrostatics, where the induced electric dipole moments of the
molecules of a dielectric material contribute to the electric
displacement field.  We explore this analogy in our analysis of
particle and fluid transport through a channel.

Our paper is organized as follows. Far-field dipolar particle response and
particle polarizability are discussed in Sec.\ \ref{Particle polarizability},
where we give a general outline of our theory.  The dipole moment of an
arbitrary induced-force distribution in a channel is evaluated in Sec.\
\ref{Dipolar moment of induced-force distribution}.  The polarizability
coefficients for a spherical particle are determined in the friction and
mobility formulations in Sec.\ \ref{Polarizability coefficients}. The results
from the preceding sections are used to analyze macroscopic suspension flow in
Secs.\ \ref{Macroscopic suspension flow} and \ref{Transport coefficients}:
Sec.\ \ref{Macroscopic suspension flow} relates the macroscopic suspension
velocity to the dipolar density per unit area of the channel, and Sec.\
\ref{Transport coefficients} provides macroscopic transport equations
describing the particle and fluid transport.  Our conclusions are drawn in
Sec.\ \ref{Conclusions}.

An important additional result of this study is the derivation of explicit
Ewald-summation expressions for the periodic Green's functions for Stokes flow
in the parallel-wall geometry.  These formulations are used in our current
numerical simulations to supplement the theoretical analysis, and our
expressions are also applicable in Stokesian-dynamics and boundary-integral
algorithms for dispersion flows in parallel-wall channels.

\section{Particle polarizability}
\label{Particle polarizability}

In this section we focus on a general discussion of the far-field
response of a spherical particle in a parallel-wall channel to an
applied lateral force $\appliedForce$, torque $\appliedTorque$, and
Poiseuille flow $\externalVelocity$ driven by a constant lateral
pressure gradient $\appliedPressureGradient$.  This response is
represented in terms of particle polarizability coefficients,
which are subsequently used to determine the macroscopic constitutive
relations describing suspension transport in a channel.

Our results for the particle response to external forcing can also be
used to construct a simplified description of particle dynamics in
parallel-wall channels.  In this simplified formulation (summarized in
Sec.\ \ref{single-scattering approximation}), the interparticle
hydrodynamic interactions are incorporated solely through the
far-field flow, in analogy to the point-particle approximation for
unbounded systems.  As discussed in \
\cite{Baron-Blawzdziewicz-Wajnryb:2008}, such a single-scattering
approximation yields accurate results if the interparticle separation
is sufficiently large.

\subsection{System definition}

We consider the dynamics of a spherical particle of radius $a$ (or an
array of such particles) in a parallel-wall channel of width $H$,
under creeping-flow conditions.  The walls are in the planes $z=0$ and
$z=H$, and the particle center is at the axis $z$ at a distance $z=Z$
from the lower wall.  The fluid velocity field satisfies the no-slip
boundary conditions on the walls and the particle surface.

For simplicity, the analysis presented in Secs.\ \ref{Particle
polarizability}--\ref{Polarizability coefficients} is for a single
particle in the channel.  However, our theoretical formulation can be
readily generalized to multiparticle systems.  We use such a
generalization in Secs.\ \ref{Macroscopic suspension flow} and
\ref{Transport coefficients}, where we consider macroscopic response
of suspension to external forcing.

\subsection{Far-field scattered flow}

In the near-field regime $\rho\sim H$ (where $\rho=|\brho|$, and
$\brho=x\ex+y\ey$ is the lateral position with respect to the particle center)
the flow field scattered by the particle, $\scatteredVelocityPart$, has a
complex 3D form that involves multiple image singularities
\cite{Bhattacharya-Blawzdziewicz:2002}.  However, in the far-field regime
($\rho\gg H$) the scattered flow tends exponentially (on the lengthscale $H$)
to a much simpler 2D Hele--Shaw flow of the form
\cite{Bhattacharya-Blawzdziewicz-Wajnryb:2006}
\begin{equation}
\label{Hele-Shaw flow}
\scatteredVelocityPart(\br)
   =-\halff\eta^{-1}z(H-z)\bnablaLat \scatteredPressurePart(\brho),
\end{equation}
where $\eta$ is the fluid viscosity, $\scatteredPressurePart$ is the
perturbation pressure, $\bnablaLat$ is the gradient operator with
respect to the lateral coordinates $\brho$, and $\br=\brho+z\ez$.

Away from the singularity at $\brho=0$, the pressure field
$\scatteredPressurePart$ satisfies the 2D Laplace equation
\begin{equation}
\label{2D Laplace equation for the pressure}
\nablaLat^2\scatteredPressurePart(\brho)=0,\qquad \brho\not=0,
\end{equation}
owing to the flow incompressibility.  For a single particle moving under the
action of an external force, torque, or Poiseuille flow, the pressure
$\scatteredPressurePart$ assumes the form of a 2D dipolar field, due to the
cylindrical symmetry of the problem and the vectorial character of the
forcing.  Accordingly, we have
\begin{equation}
\label{pressure dipole}
\scatteredPressurePart(\brho)
   =-\frac{1}{2\pi}\particleDipoleMoment\bcdot\bnablaLat\Coulomb(\brho)
   =\frac{1}{2\pi}\particleDipoleMoment\bcdot\frac{\brho}{\rho^2},
\end{equation}
where
\begin{equation}
\label{2D Coulomb potential}
\Coulomb(\brho)=-\ln(\rho)
\end{equation}
is the solution of the 2D Poisson equation 
\begin{equation}
\label{Poisson equation}
\nablaLat^2\Coulomb(\brho)=-2\pi\delta(\brho),
\end{equation}
and $\particleDipoleMoment$ is the dipole moment of the perturbation
pressure $\scatteredPressurePart$.

As discussed in the introduction, the Hele--Shaw flow \refeq{Hele-Shaw flow}
with the dipolar pressure distribution \refeq{pressure dipole} involves the
backflow effect.  Namely, on the symmetry axis parallel to the dipole moment
$\particleDipoleMoment$, the fluid displaced by the particle is moving in the
direction of $\particleDipoleMoment$, whereas on the transverse axis it is
moving in the opposite direction.

\subsection{Polarizability and mobility relations}

Since the dynamics of the system is governed by the linear Stokes
equations, the dipole moment $\particleDipoleMoment$ is linear in the
strength of the external forcing,
\begin{equation}
\label{polarizability relation}
\smallfrac{1}{12}\eta^{-1}H^3\particleDipoleMoment
   =\mobilityMatrixPT\bcdot\appliedForce
   +\mobilityMatrixPR\bcdot\appliedTorque
   +\mobilityMatrixPP\appliedPressureGradient
\end{equation}
(where the factor $\smallfrac{1}{12}\eta^{-1}H^3$ is introduced to
ensure the Lorentz symmetry for the matrix of transport coefficient,
as explained in Sec.\ \ref{Polarizability coefficients}).  The dipolar
far-field perturbation pressure \refeq{pressure dipole} is analogous
to a 2D electrostatic dipolar potential.  Relation
\refeq{polarizability relation} will thus be termed a polarizability
relation, and the linear transport coefficients
$\mobilityMatrix^{\pressure A}$ ($A=\pressure,\transl,\rot$) will be
referred to as the polarizability coefficients (by analogy with an
electrostatic problem of polarizable particles in an external electric
field). 

Apart from producing the far-field response described by Eqs.\
\refeq{Hele-Shaw flow}--\refeq{polarizability relation}, a particle in
a channel also undergoes translational and rotational motion with the
linear and angular velocities $\particleVelocity$ and
$\particleAngularVelocity$.  This rigid-body particle motion is
characterized by the standard mobility relations
\begin{subequations}
\label{standard mobility relations}
\begin{equation}
\label{translational mobility relation}
\particleVelocity = \mobilityMatrixTT\bcdot\appliedForce
+\mobilityMatrixTR\bcdot\appliedTorque -
\mobilityMatrixTP\bcdot\appliedPressureGradient,
\end{equation}
\begin{equation}
\label{rotational mobility relation}
\particleAngularVelocity = \mobilityMatrixRT\bcdot\appliedForce
+\mobilityMatrixRR\bcdot\appliedTorque -
\mobilityMatrixRP\bcdot\appliedPressureGradient,
\end{equation}
\end{subequations}
where $\mobilityMatrix^{AB}$ denotes hydrodynamic mobility coefficients.

By a partial inversion of equations \refeq{polarizability relation} and
\refeq{standard mobility relations}, the particle response can also be
expressed in a friction-relation form, where the velocities
$\particleVelocity$ and $\particleAngularVelocity$ are the independent
variables, and the forces $\appliedForce$ and $\appliedTorque$ are the
dependent quantities (cf.\ Sec.\ \ref{Polarizability coefficients}).  

Equations \refeq{polarizability relation} and \refeq{standard mobility
relations} are related.  Namely, as shown in Sec.\ \ref{Polarizability
coefficients}, the matrix of mobility and polarizability coefficients
$\mobilityMatrix^{AB}$ satisfies the Lorentz symmetry
\begin{equation}
\label{Lorentz symmetry of generalized mobility matrix}
\mobilityMatrix^{AB}=\mobilityMatrix^{BA\,\dagger},\qquad
A,B=\transl,\rot,\pressure,
\end{equation}
where the dagger denotes the transpose of a tensor \cite{scalar_coefficients}.
Equation \refeq{Lorentz symmetry of generalized mobility matrix} has
significant consequences for suspension dynamics in parallel-wall channels,
because it implies the corresponding Onsager reciprocal relation for the
matrix of kinetic coefficients in the linear constitutive relation describing
macroscopic suspension transport.

\subsubsection{Macroscopic suspension flow}
\label{subsection on Macroscopic suspension flow}

The polarizability relation \refeq{polarizability relation} (and its
multiparticle generalization) is essential for theoretical
understanding of the macroscopic behavior of suspensions confined in a
parallel-wall channel.  This relation is also important for numerical
evaluation of effective transport coefficients governing the
suspension flux.

In the following sections we demonstrate that the average suspension
velocity $\averageSuspensionVelocity$ can be expressed in terms of the
average dipole moment of the particles $\averageParticleDipoleMoment$.
Specifically, in Sec.\ \ref{Macroscopic suspension flow} it is shown
that
\begin{equation}
\label{dipolar expression for average suspension velocity}
\averageSuspensionVelocity=\channelPermeability
\left(
   -\macroscopicPressureGradient
   +\suspensionDensityA\averageParticleDipoleMoment
\right),
\end{equation}
where 
\begin{equation}
\label{channel permeability}
\channelPermeability=\smallfrac{1}{12}\eta^{-1}H^2
\end{equation}
is the permeability of a particle-free channel,
$\macroscopicPressureGradient$ is the macroscopic pressure gradient,
and $\suspensionDensityA$ is the particle number density per unit area
of the channel wall.  Equation \refeq{dipolar expression for average
suspension velocity}, supplemented with the polarizability relation
\refeq{polarizability relation} (averaged over the particle
distribution) describes the macroscopic suspension flow (cf.\ Secs.\
\ref{Macroscopic suspension flow} and \ref{Transport coefficients}).

An intuition regarding physical interpretation of Eq.\ \refeq{dipolar
expression for average suspension velocity} can be gained by
considering its electrostatic analogy.  A comparison of the
hydrodynamic and electrostatic problems indicates that the macroscopic
pressure field is analogous to the electrostatic potential, and the
pressure gradient corresponds to the electric field.  The macroscopic
suspension velocity $\averageSuspensionVelocity$ (which is divergence
free due to the fluid incompressibility) corresponds to the
electric-displacement field in the absence of external charges.
Relation \refeq{dipolar expression for average suspension velocity} is
thus similar to the expression for the electrostatic displacement
field in terms of the electric field and the induced-dipole-moment
density \cite{Jackson:1999}.  In Sec.\ \ref{Macroscopic suspension
flow} we further explore this analogy.

A macroscopic theory based on the transport equations
\refeq{polarizability relation}, \refeq{standard mobility relations},
and \refeq{dipolar expression for average suspension velocity} is
capable of describing complex phenomena that occur in suspension flows
in a parallel-wall channels.  In particular, we have shown in our
recent study \cite{Baron-Blawzdziewicz-Wajnryb:2008} that such a
theory predicts a fingering instability in evolving 2D
particle arrays (which has been confirmed by direct numerical
simulations).

\subsubsection{Single-scattering approximation}
\label{single-scattering approximation}

The amplitude $\particleDipoleMoment$ of the dipolar Hele--Shaw flow
scattered by a particle in a channel is a key quantity in a
single-scattering approximation, where the one-particle polarizability
and mobility relations \refeq{polarizability relation} and
\refeq{standard mobility relations} are combined with an assumption
that the incident flow acting on a given particle is a superposition
of the external flow and the far-field dipolar flows \refeq{Hele-Shaw
flow} produced by other particles.

The single-scattering approximation describes particle dynamics in
dilute suspensions in parallel-wall channels under the
strong-confinement condition $H\sim2a$.  Many fundamental collective
phenomena in confined suspension flows (e.g.\ instabilities of
confined particle jets \cite{Alvarez-Clement-Soto:2006}, propagation
of particle-displacement waves in 1D arrays of drops
\cite{Beatus-Tlusty-Bar_Ziv:2006}, and pattern formation in 2D regular
particle arrays \cite{Baron-Blawzdziewicz-Wajnryb:2008}) are driven by
the far-field interparticle interactions associated with the dipolar
scattered flow \refeq{Hele-Shaw flow} and \refeq{pressure dipole}.
Not only can essential qualitative features of such phenomena be
captured using the single-scattering approximation, but it also gives
accurate quantitative results if the interparticle distances are
sufficiently large \cite{point-particie-approximation}.

Quantitative predictions using the single-scattering approach require
predetermination of the polarizability and mobility coefficients in Eqs.\
\refeq{polarizability relation} and \refeq{standard mobility relations}.  We
evaluate only the polarizability coefficients $\mobilityMatrix^{\pressure B}$
($B=\transl,\rot,\pressure$), because the mobility coefficients
$\mobilityMatrix^{AB}$ ($A=\transl,\rot\ ,B=\transl,\rot,\pressure$) for
spherical particles between two parallel walls have already been calculated
\cite{
Staben-Zinchenko-Davis:2003, Jones:2004,
Bhattacharya-Blawzdziewicz-Wajnryb:2005,
Bhattacharya-Blawzdziewicz-Wajnryb:2005a,
Bhattacharya-Blawzdziewicz-Wajnryb:2006%
}.  

\section{Dipole moment of induced-force distribution}
\label{Dipolar moment of induced-force distribution}

To evaluate the induced dipole moment of a particle in a channel, we
need to solve the corresponding Stokes-flow problem.  For this
purpose, we apply the induced-force formulation and
multipolar-expansion techniques.  In this section we use the
asymptotic Liron--Mochon formula for the Green's function for Stokes
flow in the parallel-wall geometry \cite{Liron-Mochon:1976}, to relate
the dipole moment $\particleDipoleMoment$ to the force distribution
induced on the particle.  We employ these results in Sec.\
\ref{Polarizability coefficients} to calculate the polarizability
coefficients, and in Secs.\ \ref{Macroscopic suspension flow} and
\ref{Transport coefficients} to determine constitutive relations
describing macroscopic suspension flow.

\subsection{Induced-force formulation}
\label{Induced-force formulation}

In our approach, the effect of a particle on the surrounding fluid is
represented in terms of the induced-force distribution on the
particle surface
\begin{equation}
\label{induced forces}
\bF(\br)=a^{-2}\delta(r_1-a)\bff_1(\br_1),
\end{equation}
where $\br_1=\br-Z\ez$ denotes the position with respect to the
particle center, and $r_1=|\br_1|$.  By definition of the induced
force, the flow and pressure fields produced by the distribution
\refeq{induced forces} are identical to the velocity field $\bv(\br)$
and pressure $p(\br)$ in the particle presence
\cite{Cox-Brenner:1967,Mazur-Bedeaux:1974,Felderhof:1976b}.

For a particle in an external flow $\externalVelocity(\br)$, the
velocity and pressure fields can be represented by the boundary
integrals
\begin{subequations}
\label{boundary integrals for velocity and pressure fields}
\begin{equation}
\label{flow field produced by induced forces}
\bv(\br)=\externalVelocity(\br)+
  \int\GreenT(\br,\br')\bcdot\bF(\br')\diff\br',
\end{equation}
\begin{equation}
\label{pressure produced by induced forces}
p(\br)=\externalPressure(\br)+
  \int\GreenQ(\br,\br')\bcdot\bF(\br')\diff\br',
\end{equation}
\end{subequations}
where $\externalPressure(\br)$ is the external pressure associated
with the flow $\externalVelocity(\br)$, and $\GreenT$ and $\GreenQ$
are the velocity and pressure Green's functions for a parallel-wall
channel.

The Green's functions for Stokes flow in the parallel-wall geometry
were investigated using several different methods
\cite{%
Liron-Mochon:1976,%
Bhattacharya-Blawzdziewicz:2002,%
Staben-Zinchenko-Davis:2003,%
Jones:2004,%
Bhattacharya-Blawzdziewicz-Wajnryb:2005,%
Bhattacharya-Blawzdziewicz-Wajnryb:2005a,%
Bhattacharya-Blawzdziewicz-Wajnryb:2006%
}.
In the Cartesian-representation approach proposed by our group
\cite{%
Bhattacharya-Blawzdziewicz-Wajnryb:2005,%
Bhattacharya-Blawzdziewicz-Wajnryb:2005a,%
Bhattacharya-Blawzdziewicz-Wajnryb:2006%
}, 
the velocity and pressure Green's functions $\GreenT(\br,\br')$ and
$\GreenQ(\br,\br')$ are represented in terms of lateral Fourier
integrals of simple matrix products.  The explicit formulas are listed
in Appendix \ref{Expressions for velocity and pressure Green's
functions}.

For a particle moving with the translational and angular velocities
$\particleVelocity$ and $\particleAngularVelocity$, the flow field
\refeq{flow field produced by induced forces}, evaluated at the
particle surface $S$, equals the rigid body velocity of the particle
\begin{equation}
\label{rigid-body velocity of drop}
\bv(\br)=\bv^{\rm rb}(\br)\equiv\bU+\bOmega\btimes\br_1,\qquad \br\in S.
\end{equation}
With the above boundary condition, Eq.\ \refeq{flow field produced by
induced forces} yields a boundary-integral equation for the induced
forces.

For a given induced-force distribution \refeq{induced forces} the
force and torque acting on the particle can be evaluated using
expressions
\begin{equation}
\label{force and torque}
   \appliedForce=\int\bF(\br)\diff\br,
\qquad
   \appliedTorque=\int\br\btimes\bF(\br)\diff\br.
\end{equation}
In the following section we determine the corresponding relation for
the amplitude of the dipolar far-field flow produced by the particle.

\subsection{Dipole moment}

The dipolar strength $\particleDipoleMoment$ of a force distribution
$\bF$ can be obtained from Eqs.\ \refeq{boundary integrals for
velocity and pressure fields} using asymptotic far-field expressions
for the Green's function $\GreenT$ and $\GreenQ$.  (Such expressions
were first derived by Liron and Mochon \cite{Liron-Mochon:1976}; an
alternative and much simpler derivation via a lubrication expansion is
given in \cite{Bhattacharya-Blawzdziewicz:2008}).

As discussed in \cite{Bhattacharya-Blawzdziewicz-Wajnryb:2006}, the asymptotic
form of the flow and pressure Green's functions, $\GreenTasymptotic(\br,\br')$
and $\GreenQasymptotic(\br,\br')$, can be represented by the following
formulas
\begin{subequations}
\label{non-periodic asymptotic Green's functions}
\begin{equation}
\label{non-periodic asymptotic Green's function T}
\GreenTasymptotic(\br,\br')
   =-\halff\eta^{-1} z(H-z)\bnabla\GreenQasymptotic(\br,\br'),
\end{equation}
\begin{equation}
\label{non-periodic asymptotic Green's function Q}
\GreenQasymptotic(\br,\br')=-3\pi^{-1} H^{-3}
        \bnabla\Coulomb(\brho-\brho')z'(H-z'),
\end{equation}
\end{subequations}
where $\brho=x\ex+y\ey$ and $\brho'=x'\ex+y'\ey$ are the lateral
position vectors, and $\Coulomb(\brho)$ is the point-source solution
\refeq{2D Coulomb potential} of the 2D Poisson equation
\refeq{Poisson equation}.  We note that the asymptotic Hele--Shaw
flow and pressure fields \refeq{non-periodic asymptotic Green's
functions} satisfy the Stokes equations exactly.  However, these
fields have a different singularity than the original point-force
singularity of the full Green's functions $\GreenT$ and $\GreenQ$.  

According to Eqs.\ \refeq{non-periodic asymptotic Green's function Q} and
\refeq{2D Coulomb potential}, the far-field pressure
\begin{equation}
\label{far field pressure produced by point force}
\dipolarPressure(\brho)=\GreenQasymptotic(\br,\br')\bcdot\lateralPointForce
\end{equation}
produced by a lateral point force $\lateralPointForce$ has a form of a
2D potential dipole 
\begin{equation}
\label{pressure dipole produced by point force}
\dipolarPressure(\brho)
   =-\frac{1}{2\pi}\pressureDipoleMoment
     \bcdot\bnablaLat\Coulomb(\brho-\brho')
\end{equation}
with the dipole moment
\begin{equation}
\label{dipolar moment pressure produced by point force}
\pressureDipoleMoment=
  6H^{-3}z'(H-z')\lateralPointForce.
\end{equation}
The streamlines of the corresponding velocity field
\begin{equation}
\label{far-field flow produced by point force}
\dipolarFlow(\br)=-\halff\eta^{-1}z(H-z)\bnablaLat\dipolarPressure(\brho)
\end{equation}
also follow a 2D dipolar pattern.  The quadratic dependence of the dipolar
strength \refeq{dipolar moment pressure produced by point force} on the
position $z'$ of the point where the force is applied follows from the
Lorentz's symmetry of the Green's function \refeq{non-periodic asymptotic
Green's function T} \cite{Bhattacharya-Blawzdziewicz-Wajnryb:2006}.  We note
that the flow field produced by a transverse force (i.e.\ a force pointing in
the $z$ direction) is exponentially small in the far-field domain.

Similar to the corresponding electrostatic problem, the dipole moment
\refeq{dipolar moment pressure produced by point force} represents not only
the amplitude of the far-field pressure \refeq{far field pressure produced by
point force} but also the strength of the dipolar-pressure source,
\begin{equation}
\label{Poisson equation for dipolar pressure}
\nablaLat^2\dipolarPressure(\brho)
   =\pressureDipoleMoment\bcdot\bnablaLat\delta(\brho-\brho').
\end{equation}
In Sec.\ \ref{Macroscopic suspension flow} the above expression will be used
in our derivation of the relation between the macroscopic dipolar-strength
density per unit area of the channel and the average suspension flow.

The dipole moment of the far-field perturbation pressure
\refeq{pressure dipole} produced by a particle in a channel is
evaluated by integrating \refeq{dipolar moment pressure produced by
point force} over the particle surface.  Taking
$\lateralPointForce=\lateralUnitTensor\bF(\br)$, where 
\begin{equation}
\label{lateral unit tensor}
\lateralUnitTensor=\ex\ex+\ey\ey
\end{equation}
denotes the projection operator onto the lateral directions $x$ and
$y$, and \bF(\br) is the induced-force distribution \refeq{induced
forces}, we get
\begin{equation}
\label{expression for dipolar moment of induced force distribution}
\dipoleMoment
   =6H^{-3}\int z'(H-z')
      \lateralUnitTensor\bcdot\bF(\br')\diff\br'.
\end{equation}
In the following section, Eq.\ \refeq{expression for dipolar moment of
induced force distribution} is used to determine the polarizability
coefficients of a particle.

\section{Polarizability coefficients}
\label{Polarizability coefficients}

\subsection{Multipolar expansion}
\label{Multipolar expansion}

To determine the force distribution induced on the particle surface,
the boundary-value problem \refeq{flow field produced by induced
forces} and \refeq{rigid-body velocity of drop} is solved using the
multipolar-expansion technique
\cite{%
Bhattacharya-Blawzdziewicz-Wajnryb:2005,%
Bhattacharya-Blawzdziewicz-Wajnryb:2005a,%
Bhattacharya-Blawzdziewicz-Wajnryb:2006%
}.
In our method, the induced-force distribution and the flow field in the
system are expanded into the conjugate sets of basis functions
introduced in
\cite{%
Cichocki-Felderhof-Schmitz:1988%
}.  
In particular, we have the expansions
\begin{subequations}
\label{multipolar expansions}
\begin{equation}
\label{induced force in terms of multipoles}
\bF(\br)
   =\sum_{lm\sigma}
      f(lm\sigma)
         \reciprocalSphericalBasisP{lm\sigma}(\br_1),
\end{equation}
\begin{equation}
\label{expansion of external flow}
\bv^{\rm rb}(\br)-\externalVelocity(\br)
   =\sum_{lm\sigma}c(lm\sigma)\sphericalBasisP{lm\sigma}(\br_1),
\end{equation}
\end{subequations}
where the left-hand side of Eq.\ \refeq{expansion of external flow}
describes the external flow with respect to the rigid-body particle
motion \refeq{rigid-body velocity of drop}.

In the above relations $\reciprocalSphericalBasisP{lm\sigma}$ and
$\sphericalBasisP{lm\sigma}$ are the conjugate spherical basis
functions associated with the non-singular solutions of Stokes
equations in spherical coordinates
\cite{%
Bhattacharya-Blawzdziewicz-Wajnryb:2005,%
Bhattacharya-Blawzdziewicz-Wajnryb:2005a,%
Cichocki-Felderhof-Schmitz:1988%
},
and $f(lm\sigma)$ and $c(lm\sigma)$ are the corresponding expansion
coefficients \cite{no_delta}.  The flow $\scatteredVelocityPart$
scattered by the particle has a similar expansion in terms of singular
basis functions $\sphericalBasisM{lm\sigma}$.  Here $l=1,2,\ldots$ and
$m=0,\pm1,\ldots,\pm l$ are the spherical harmonic orders of the basis
functions, and the index $\sigma=0,1,2$ corresponds to the three types
of Lamb's solutions for Stokes flow.

By inserting expansions \refeq{multipolar expansions} into relation
\refeq{flow field produced by induced forces} evaluated at the particle
surface one gets a linear algebraic equation of the form
\cite{%
Bhattacharya-Blawzdziewicz-Wajnryb:2005,%
Bhattacharya-Blawzdziewicz-Wajnryb:2005a%
}
\begin{equation}
\label{induced force equations}
   \sum_{l'm'\sigma'}
      \GreenWallTotElement(lm\sigma\mid l'm'\sigma')
      f(l'm'\sigma')
      =c(lm\sigma),
\end{equation}
where the matrix elements $\GreenWallTotElement(lm\sigma\mid
l'm'\sigma')$ are defined in terms of the multipolar projections of
the  Green's function $\GreenT$,
\begin{equation}
\label{matrix elements}
\GreenWallTotElement(lm\sigma\mid l'm'\sigma')
        =\langle\reciprocalSphericalBasisP{lm\sigma}(\br_1)
\mid
   \GreenT
\mid
   \reciprocalSphericalBasisP{l'm'\sigma'}(\br_1)\rangle.
\end{equation}
In Eq.\ \refeq{matrix elements} we use the Dirac's bra--ket notation
\cite{Shankar:2008}
\begin{equation}
\label{bra--ket ntation}
\langle \bA\mid\bB\rangle=\int \bA^*(\br)\bcdot\bB(\br)\diff\br,
\end{equation}
with the asterisk denoting the complex conjugate.  

Equation \refeq{matrix elements} is valid for a particle with the
no-slip boundary conditions on its surface.  For other boundary
conditions (e.g., spherical viscous drops) the diagonal elements of
\refeq{matrix elements} need to be modified to include an appropriate
single-particle scattering matrix
\cite{%
Bhattacharya-Blawzdziewicz-Wajnryb:2005,%
Bhattacharya-Blawzdziewicz-Wajnryb:2005a,%
Bhattacharya-Blawzdziewicz-Wajnryb:2006%
}.
It should also be noted that equation \refeq{induced force equations} is given
here for a single particle, but general expressions for a multiparticle system
are readily available
\cite{%
Bhattacharya-Blawzdziewicz-Wajnryb:2005,%
Bhattacharya-Blawzdziewicz-Wajnryb:2005a,%
Bhattacharya-Blawzdziewicz-Wajnryb:2006%
}.

\subsection{Generalized friction matrix}
\label{Generalized friction matrix}

The response of the system to a given rigid-body particle motion
\refeq{rigid-body velocity of drop} and external parabolic flow
\begin{equation}
\label{external parabolic flow}
\externalVelocity=-\halff\eta^{-1}z(H-z)\bnabla\externalPressure
\end{equation}
driven by a constant lateral pressure gradient
$\bnabla\externalPressure$ can be characterized by the generalized
resistance relation 
\begin{equation}
\label{generalized resistance relation}
\left[
   \begin{array}{c}
      \totForce\\
      \totTorque\\
      \smallfrac{1}{12}H^3\dipoleMoment
   \end{array}
\right]
=
\left[
   \begin{array}{ccc}
      \resistanceMatrixTT&\resistanceMatrixTR&\resistanceMatrixTP\\
      \resistanceMatrixRT&\resistanceMatrixRR&\resistanceMatrixRP\\
      \resistanceMatrixPT&\resistanceMatrixPR&\resistanceMatrixPP\\
   \end{array}
\right]
   \bcdot
\left[
   \begin{array}{c}
      \particleVelocity\\
      \particleAngularVelocity\\
      \eta^{-1}\bnabla\externalPressure
   \end{array}
\right].
\end{equation}
As discussed below, the factor $\smallfrac{1}{12}H^3$ that multiplies
the dipole moment $\dipoleMoment$ is needed to ensure the Lorentz
symmetry \refeq{Lorentz symmetry of generalized friction matrix} of
the resistance matrix $\resistanceMatrix$.  The coefficients in the
bottom row of the matrix $\resistanceMatrix$ characterize the
particle polarizability.

The generalized resistance tensors $\resistanceMatrix^{AB}$ can be
determined by solving Eq.\ \refeq{induced force equations} to evaluate
the induced-force amplitudes $f(lm\sigma)$ in terms of the flow
expansion coefficients $c(lm\sigma)$.  The solution can be represented
by the matrix relation
\begin{equation}
\label{solution for induced forces}
f(lm\sigma)=\sum_{l'm'\sigma'}
            \GrandFrictionElement(lm\sigma\mid l'm'\sigma')c(l'm'\sigma'),
\end{equation}
where $\GrandFrictionElement(lm\sigma\mid l'm'\sigma')$ denotes the
elements of the matrix inverse to the Green's matrix
$\GreenWallTotElement(lm\sigma\mid l'm'\sigma')$.

To obtain the resistance formula \refeq{generalized resistance relation}, Eq.\
\refeq{solution for induced forces} has to be supplemented with appropriate
expressions that relate the force $\totForce$, torque $\totTorque$, and dipole
moment $\dipoleMoment$ to the expansion coefficients $f(lm\sigma)$ of the
induced force distribution \refeq{induced force in terms of multipoles}.  We
also need the relations for the expansion coefficients $c(lm\sigma)$
associated with a given rigid-body particle motion \refeq{rigid-body velocity
of drop} and external flow \refeq{external parabolic flow}.  These relations
can be expressed in terms of the transformation vectors
$\frictionProjectionVector$, discussed in Appendix \ref{Projection vectors X}.
Namely, we have
\begin{subequations}
\label{multipolar expressions for F T D}
\begin{equation}
\label{multipolar expression for F}
\totForce=\sum_{lm\sigma}
         \frictionProjectionVector(\transl\mid lm\sigma)f(lm\sigma),
\end{equation}

\begin{equation}
\label{multipolar expression for T}
\totTorque=\sum_{lm\sigma}
          \frictionProjectionVector(\rot\mid lm\sigma)f(lm\sigma),
\end{equation}
\begin{equation}
\label{multipolar expression for D}
\smallfrac{1}{12}H^3\dipoleMoment=\sum_{lm\sigma}
          \frictionProjectionVector(\pressure\mid lm\sigma)f(lm\sigma),
\end{equation}
\end{subequations}
and
\begin{eqnarray}
\label{coefficients c for given particle motion and parabolic flow}
   c(lm\sigma)
      &=&\frictionProjectionVector(lm\sigma\mid\transl)\bcdot\particleVelocity
      +\frictionProjectionVector(lm\sigma\mid\rot)
                         \bcdot\particleAngularVelocity
\nonumber\\
   &+&\frictionProjectionVector(lm\sigma\mid\pressure)
                      \bcdot\eta^{-1}\bnabla\externalPressure,
\end{eqnarray}
where the transformation vectors $\frictionProjectionVector$ satisfy
the symmetry relation
\begin{equation}
\label{symmetry of vectors X}
\frictionProjectionVector(lm\sigma\mid A)
   =\frictionProjectionVector^*(A\mid lm\sigma).
\end{equation}
Explicit expressions for the transformation vectors \refeq{symmetry of
vectors X} are listed in Appendix \ref{Projection vectors X}.  We note
that only several matrix elements with small values of the indices
$l,m,\sigma$ are nonzero, according to Eqs.\ \refeq{force
transformation vectors} and \refeq{condition for nonzero elements of C
matrix}.

Combining Eqs.\ \refeq{solution for induced
forces}--\refeq{coefficients c for given particle motion and parabolic
flow} yields the relation
\begin{widetext}
\begin{equation}
\label{expression for resistance coefficients}
\resistanceMatrix^{AB}
   =\sum_{lm\sigma}\sum_{l'm'\sigma'}
    \frictionProjectionVector(A\mid lm\sigma)
    \GrandFrictionElement(lm\sigma\mid l'm'\sigma')
    \frictionProjectionVector(l'm'\sigma'\mid B),
\qquad A,B=\transl,\rot,\pressure,
\end{equation}
\end{widetext}
for the translational, rotational, and polarizability components of
the generalized friction matrix.  Taking into account relation
\refeq{symmetry of vectors X} and the Lorentz symmetry of the matrix
$\GrandFrictionElement$
\cite{%
Bhattacharya-Blawzdziewicz-Wajnryb:2005a,
Bhattacharya-Blawzdziewicz-Wajnryb:2005%
},
\begin{equation}
\label{Lorentz symmetry of grad friction matirx}
\GrandFrictionElement(lm\sigma\mid l'm'\sigma')
   =\GrandFrictionElement^*(l'm'\sigma'\mid lm\sigma),
\end{equation}
we obtain the corresponding symmetry of the generalized friction matrix
\begin{equation}
\label{Lorentz symmetry of generalized friction matrix}
\resistanceMatrix^{AB}=\resistanceMatrix^{BA\,\dagger}.
\end{equation}

The symmetry relation \refeq{Lorentz symmetry of generalized friction
matrix} with $A,B=\transl,\rot$ corresponds to the well-known Lorentz
symmetry of the standard resistance matrix \cite{Kim-Karrila:1991}.
For $A=\pressure$ or $B=\pressure$, however, the symmetry relation
\refeq{Lorentz symmetry of generalized friction matrix} is new, and
its importance becomes clear in Sec.\ \ref{Macroscopic suspension
flow}, where we explain the relationship between the dipole moment
induced on the particles and the macroscopic suspension flow.

\subsection{Mobility formulation}
\label{Mobility formulation}

The polarizability and mobility relations \refeq{polarizability
relation} and \refeq{standard mobility relations} can be obtained by
partially inverting the generalized resistance relation
\refeq{generalized resistance relation}.  In the matrix representation
analogous to \refeq{generalized resistance relation} we have
\begin{equation}
\label{generalized mobility relation}
\left[
   \begin{array}{c}
      \particleVelocity\\
      \particleAngularVelocity\\
      \smallfrac{1}{12}\eta^{-1} H^3\dipoleMoment
   \end{array}
\right]
=
\left[
   \begin{array}{ccr}
      \mobilityMatrixTT&\mobilityMatrixTR&\mobilityMatrixTP\\
      \mobilityMatrixRT&\mobilityMatrixRR&\mobilityMatrixRP\\
      \mobilityMatrixPT&\mobilityMatrixPR&-\mobilityMatrixPP\\
   \end{array}
\right]
   \bcdot
\left[
   \begin{array}{c}
      \totForce\\
      \totTorque\\
      -\bnabla\externalPressure
   \end{array}
\right],
\end{equation}
where 
\begin{equation}
\label{mobility matrix as inverse of friction matrix}
\left[
   \begin{array}{cc}
      \mobilityMatrixTT&\mobilityMatrixTR\\
      \mobilityMatrixRT&\mobilityMatrixRR\\
   \end{array}
\right]
=
\left[
   \begin{array}{cc}
      \resistanceMatrixTT&\resistanceMatrixTR\\
      \resistanceMatrixRT&\resistanceMatrixRR\\
   \end{array}
\right]^{-1},
\end{equation}
\begin{equation}
\label{tp mobilities}
\eta\left[
   \begin{array}{c}
      \mobilityMatrixTP\\
      \mobilityMatrixRP\\
   \end{array}
\right]
=
\left[
   \begin{array}{cc}
      \resistanceMatrixTT&\resistanceMatrixTR\\
      \resistanceMatrixRT&\resistanceMatrixRR\\
   \end{array}
\right]^{-1}
\bcdot
\left[
   \begin{array}{c}
      \resistanceMatrixTP\\
      \resistanceMatrixRP\\
   \end{array}
\right],
\end{equation}
\begin{equation}
\label{pt mobilities}
\left[
   \begin{array}{cc}
      \mobilityMatrixPT&\mobilityMatrixPR\\
   \end{array}
\right]=
\left[
   \begin{array}{c}
      \mobilityMatrixTP\\
      \mobilityMatrixRP\\
   \end{array}
\right]^\dagger,
\end{equation}
\begin{equation}
\label{pp mobilities}
\eta^2\mobilityMatrixPP=\resistanceMatrixPP
-\left[
   \begin{array}{cc}
      \resistanceMatrixPT&\resistanceMatrixPR\\
   \end{array}
\right]
\bcdot
\left[
   \begin{array}{cc}
      \resistanceMatrixTT&\resistanceMatrixTR\\
      \resistanceMatrixRT&\resistanceMatrixRR\\
   \end{array}
\right]^{-1}
\bcdot
\left[
   \begin{array}{c}
      \resistanceMatrixTP\\
      \resistanceMatrixRP\\
   \end{array}
\right].
\end{equation}
Similar to the corresponding property of the friction matrix, the
generalized mobility matrix $\mobilityMatrix^{AB}$ satisfies the
Lorentz symmetry \refeq{Lorentz symmetry of generalized mobility
matrix}, which is obtained by combining \refeq{Lorentz symmetry of
generalized friction matrix} with \refeq{mobility matrix as inverse
of friction matrix}--\refeq{pp mobilities}.  Note that the negative
sign is incorporated into the pressure-gradient term on the right-hand
side of Eq.\ \refeq{generalized mobility relation} to obtain positive
signs of the ``$\pressure\transl$'' and ``$\pressure\rot$'' components
of the matrix $\mobilityMatrix$.  We have also included factors $\eta$
and $\eta^2$ in Eqs.\ \refeq{tp mobilities} and \refeq{pp mobilities},
to ensure that the polarizability and mobility coefficients have
appropriate dimensionalities.

The generalized mobility and friction tensors \refeq{generalized
resistance relation} and \refeq{generalized mobility relation} are
given here only for a single sphere.  However, their multiparticle
generalizations can easily be obtained.

\subsection{Tensorial form}
\label{Tensorial form}

For a single spherical particle, considered herein, the friction and
mobility tensors $\generalTensor^{AB}=\resistanceMatrix^{AB}$,
$\mobilityMatrix^{AB}$ are invariant with respect to rotation around
the axis $z$.  Noting that $\totTorque$ and $\particleAngularVelocity$
are pseudo-vectors, and that $\bnabla\externalPressure$ has only
lateral components we find that
\begin{subequations}
\label{tensorial form}
\begin{equation}
\label{tensorial form tt, rr}
\generalTensor^{AA}
   =\generalTensorCoefficient^{AA}\lateralUnitTensor
   +\generalTensorCoefficient^{AA}_\perp\ez\ez,
         \qquad A=\transl,\rot,
\end{equation}
\begin{equation}
\label{tensorial form pt, pp}
\generalTensor^{AB}
=\generalTensorCoefficient^{AB}\lateralUnitTensor,
         \qquad A=\parabolic,\quad B=\transl,\parabolic,
\end{equation}
\begin{equation}
\label{tensorial form tr, pr}
\generalTensor^{AB}
=\generalTensorCoefficient^{AB}\lateralAlternatingTensor,
         \qquad A=\transl,\parabolic,\quad B=\rot,
\end{equation}
\end{subequations}
where $\lateralUnitTensor$ is the lateral unit tensor \refeq{lateral
unit tensor}, and
\begin{equation}
\label{lateral alternating tensor}
\lateralAlternatingTensor=\ex\ey-\ey\ex
\end{equation}
is the lateral alternating tensor.  The remaining components of the
tensors $\generalTensor^{AB}$ are obtained using the reciprocal
relations \refeq{Lorentz symmetry of generalized mobility matrix} and
\refeq{Lorentz symmetry of generalized friction matrix}.  In Eqs.\
\refeq{tensorial form}, the lateral transport coefficients are denoted
by $\generalTensorCoefficient^{AB}$, and the transverse coefficients
by $\generalTensorCoefficient^{AB}_\perp$.

\begin{figure}
\includegraphics*{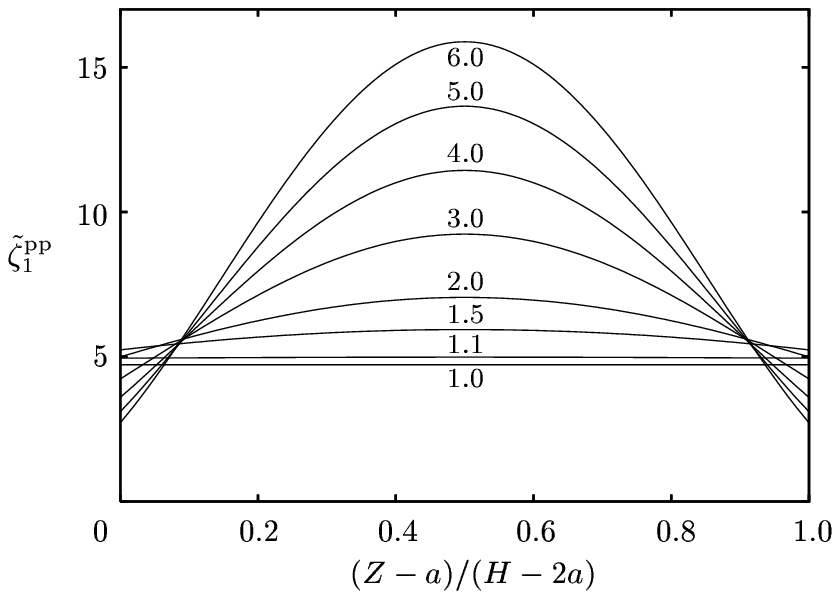}
\caption{\small
Normalized polarizability coefficient \refeq{Normalized one-particle
resistance coefficient pp}, versus particle position scaled by the
vertical space available in the channel, for channel width $H/(2a)$
(as labeled).  Coefficient $\normalizedResistanceCoefficientPP$
characterizes the amplitude of the far-field flow produced by an
immobile particle in an external pressure-driven flow.
}
\label{zeta pp one particle plot}
\end{figure}

\begin{figure}
\includegraphics*{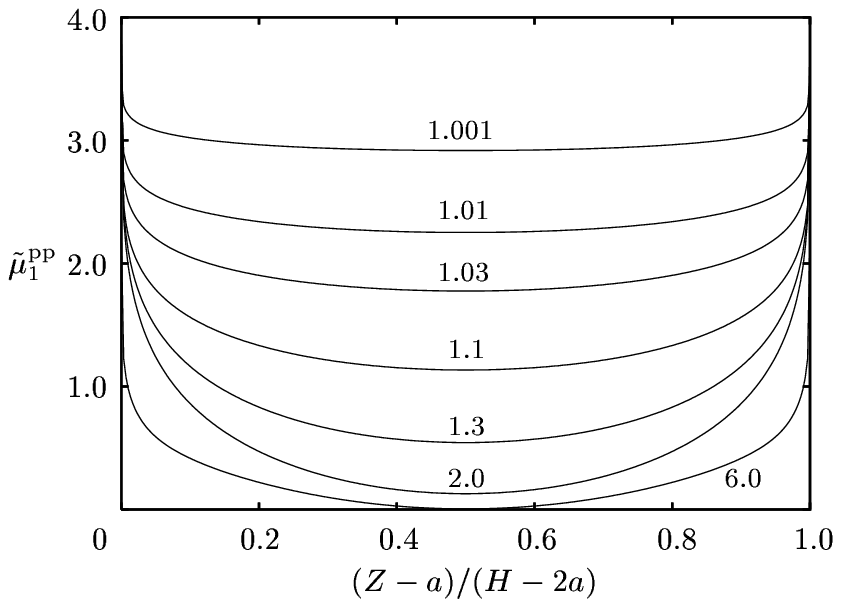}
\caption{\small
Normalized polarizability coefficient \refeq{Normalized one-particle
mobility coefficient pp}, versus particle position scaled by the
vertical space available in the channel, for channel width $H/(2a)$
(as labeled).  Coefficient $\normalizedMobilityCoefficientPP$
characterizes the amplitude of the far-field flow produced by a freely
moving particle in an external pressure-driven flow.
}
\label{mu pp one particle plot}
\end{figure}

\begin{figure}
\includegraphics*{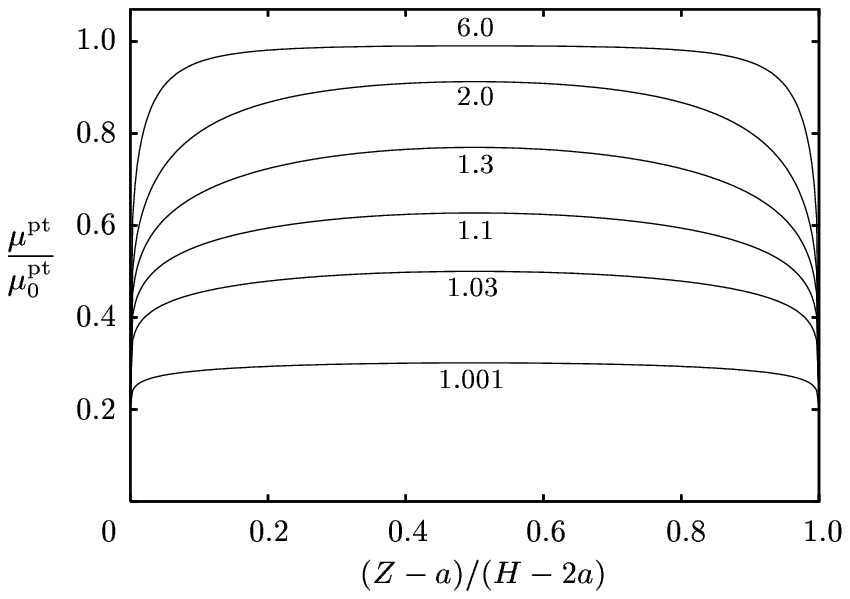}
\caption{\small
Translational polarizability coefficient $\mobilityCoefficientPT$,
normalized by the corresponding result for the point force
\refeq{point-force dipole-moment coefficient}, versus particle
position scaled by the vertical space available in the channel, for
channel width $H/(2a)$ (as labeled).  Coefficient
$\mobilityCoefficientPT$ characterizes the amplitude of the far-field
flow produced by a particle moving under the action of a lateral force.
}
\label{mu pt one particle plot}
\end{figure}

\begin{figure}
\includegraphics*{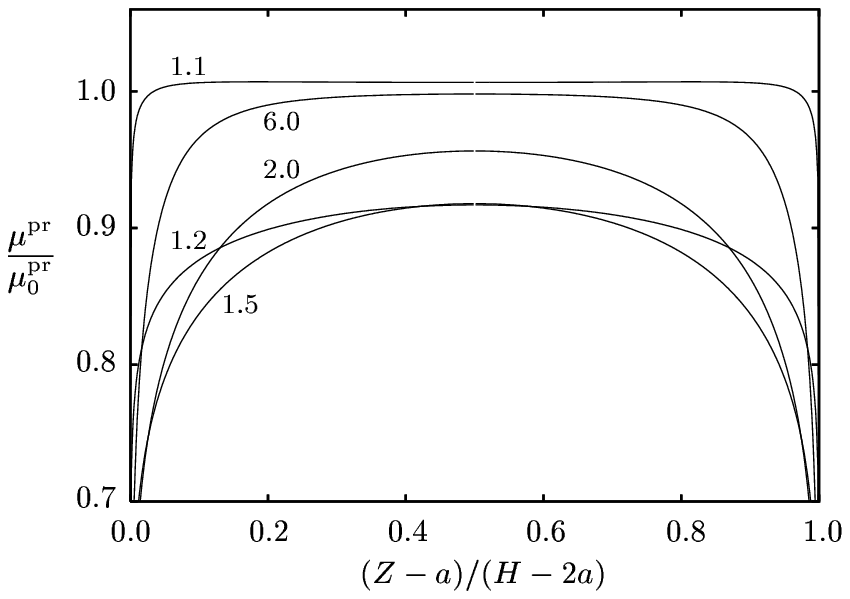}
\caption{\small
Translational polarizability coefficient $\mobilityCoefficientPR$,
normalized by the corresponding result for the point torque
\refeq{point-torque dipole-moment coefficient}, versus particle
position scaled by the vertical space available in the channel, for
channel width $H/(2a)$ (as labeled).  Coefficient
$\mobilityCoefficientPR$ characterizes the amplitude of the far-field
flow produced by a particle moving under the action of a lateral torque.
}
\label{mu pr one particle plot}
\end{figure}

\subsection{Numerical results}

In this section we present our numerical results for the generalized
resistance and mobility matrices $\resistanceMatrix$ and
$\mobilityMatrix$.  Since we are concerned here with the effect of the
far-field flow on the system dynamics, we focus on the polarizability
components that relate the external forcing to the induced dipole
moment of the particle.

We note that the translational and rotational components of the friction and
mobility matrices were calculated by Jones \cite{Jones:2004} and by our group
\cite{%
Bhattacharya-Blawzdziewicz-Wajnryb:2005,
Bhattacharya-Blawzdziewicz-Wajnryb:2005a%
}.
The effect of an external Poiseuille flow on particle motion has also
been determined
\cite{%
Jones:2004,
Staben-Zinchenko-Davis:2003,
Bhattacharya-Blawzdziewicz-Wajnryb:2006%
}.
The polarizability coefficients
$\generalTensorCoefficient^{\parabolic\transl}$ and
$\generalTensorCoefficient^{\parabolic\rot}$
($\generalTensorCoefficient=\resistanceCoefficient,
\mobilityCoefficient$) can be calculated from these earlier results by
invoking the Lorentz symmetries \refeq{Lorentz symmetry of generalized
mobility matrix} and \refeq{Lorentz symmetry of generalized friction
matrix}. The coefficient
$\generalTensorCoefficient^{\parabolic\parabolic}$, introduced here,
has never been considered.

\subsubsection{Immobile particle}

The dependence of the polarizability coefficient
$\resistanceCoefficient^{\parabolic\parabolic}$ on the channel width
and the transverse particle position $Z$ is depicted in Fig.\
\ref{zeta pp one particle plot}.  The results are presented for the
normalized coefficient
\begin{equation}
\label{Normalized one-particle resistance coefficient pp}
\normalizedResistanceCoefficientPP
   =\frac{12\resistanceCoefficientPP}{\pi\eta H^3a^2},
\end{equation}
where the subscript $\oneParticle$ indicates that the quantity
\refeq{Normalized one-particle resistance coefficient pp} is evaluated
for an isolated particle in the channel.  According to equation
\refeq{generalized resistance relation}, the coefficient
\refeq{Normalized one-particle resistance coefficient pp} represents
the dimensionless dipole moment
\begin{equation}
\label{dimensionless dipole moment}
\tilde D=\frac{|\dipoleMoment|}{\pi a^2|\bnabla\externalPressure|}
\end{equation}
produced by the imposed pressure gradient $\bnabla\externalPressure$.
As discussed in Sec.\ \ref{Fixed bed of particles}, the normalized
polarizability coefficient $\normalizedResistanceCoefficientPP$
appears as the leading-order term in the volume-fraction expansion of
the permeability of a fixed bed of particles in the channel.  The
normalization has been chosen to simplify the resulting expression.

The results in Fig.\ \ref{zeta pp one particle plot} indicate that the induced
dipole moment \refeq{dimensionless dipole moment} has a pronounced maximum for
a particle in the midplane of the channel, especially for larger values of the
channel width $H$.  This behavior results from two factors: First, the flow
acting on an immobile particle is the strongest in the center of the channel;
and second, the induced force applied at point $z'$ produces the strongest
far-field flow for $z'=H/2$, according to Eq.\ \refeq{dipolar moment pressure
produced by point force}.

\subsubsection{Suspended particle}

For a freely moving suspended particle, the numerical results are
shown for three polarizability coefficients: $\mobilityCoefficientPP$,
$\mobilityCoefficientPT$, and $\mobilityCoefficientPR$.  These
coefficients characterize the induced dipole moment produced by the
external flow, force and torque acting on the suspended particle.

Figure \ref{mu pp one particle plot} illustrates the dependence of the
polarizability coefficient $\mobilityCoefficientPP$ on the channel
width and the particle position.  The normalization of the results is
analogous to the normalization used in Fig.\ \ref{zeta pp one particle
plot} for $\resistanceCoefficientPP$,
\begin{equation}
\label{Normalized one-particle mobility coefficient pp}
\normalizedMobilityCoefficientPP
   =\frac{12\eta\mobilityCoefficientPP}{\pi H^3a^2},
\end{equation}
except that the viscosity appears in the numerator, because of the
factor $\eta$ in the definition \refeq{pp mobilities} of
$\mobilityCoefficientPP$.  The normalized polarizability coefficient
\refeq{Normalized one-particle mobility coefficient pp} is equivalent
to the dimensionless dipole moment \refeq{dimensionless dipole moment}
produced by the pressure gradient $\bnabla\externalPressure$ acting on
a force- and torque-free particle.

A comparison of the results shown in Fig.\ \ref{mu pp one particle
plot} with those depicted in Fig.\ \ref{zeta pp one particle plot}
indicates that
\begin{equation}
\label{inequality between mobility and resistance pp}
\normalizedMobilityCoefficientPP\le\normalizedResistanceCoefficientPP,
\end{equation}
where the equal sign holds for a particle touching the wall.  In such
configurations the particle cannot move, owing to the diverging
lubrication forces.  Thus the particle polarizability is the same in
the mobility and friction formulations.  Relation \refeq{inequality
between mobility and resistance pp} is consistent with a general
observation that the energy dissipation is larger in a system with
more constraints, which can be demonstrated using variational
techniques \cite{Blawzdziewicz-Wajnryb-Given-Hubbard:2005}.  Relation
\refeq{inequality between mobility and resistance pp} can also be
directly obtained from \refeq{pp mobilities}, by observing that the
translation--rotation resistance matrix is positive definite.

The results shown in Fig.\ \ref{mu pp one particle plot} indicate that
$\normalizedMobilityCoefficientPP$ has a minimum at $Z=H/2$, and that
for large values of $H/(2a)$ the minimal value nearly vanishes.  This
is because for force- and torque-free particle and large wall
separation, the dominant contribution to the dipole moment
$\dipoleMoment$ comes from the stresslet induced on the particle due
to the local velocity gradient of the imposed parabolic flow.  At the
center of the channel, the imposed velocity gradient vanishes, so
there is no stresslet contribution.

Our results for the translational and rotational polarizability
coefficients $\mobilityCoefficientPT$ and $\mobilityCoefficientPR$ are
shown in Figs.\ \ref{mu pt one particle plot} and \ref{mu pr one
particle plot}.  To emphasize the effect of the particle size and
position on the dipole moment $\dipoleMoment$, the translational
polarizability is normalized by the coefficient
\begin{equation}
\label{point-force dipole-moment coefficient}
\pointForceDipoleMomentCoefficient(Z)=\halff\eta^{-1}Z(H-Z),
\end{equation}
corresponding to the dipole moment \refeq{dipolar moment pressure
produced by point force} of a unit lateral point force (Stokeslet)
applied to the fluid at the position $Z$.  Similarly, the rotational
polarizability is normalized by the coefficient
\begin{equation}
\label{point-torque dipole-moment coefficient}
\pointTorqueDipoleMomentCoefficient(Z)=\halff\eta^{-1}(\halff H-Z),
\end{equation}
representing the dipole moment associated with the dipolar far-field
flow produced by a unit lateral point torque (rotlet)
\cite{%
Hackborn:1990,
Bhattacharya-Blawzdziewicz-Wajnryb:2006,
Bhattacharya-Blawzdziewicz:2008%
}.  Note that \refeq{point-force dipole-moment coefficient} is symmetric and
\refeq{point-torque dipole-moment coefficient} is antisymmetric with respect
to the channel center.

The results in Figs.\ \ref{mu pt one particle plot} and \ref{mu pr one
particle plot} indicate that
$\mobilityCoefficientPT/\pointForceDipoleMomentCoefficient \approx
\mobilityCoefficientPR/\pointTorqueDipoleMomentCoefficient\approx1$ for
$H/(2a)\gg1$, except for the regions adjacent to the walls, where the
normalized polarizabilities decrease logarithmically to zero for a particle
touching a wall.  The normalized polarizability
$\mobilityCoefficientPT/\pointForceDipoleMomentCoefficient$ decreases
monotonically with decreasing wall separation, whereas
$\mobilityCoefficientPR/\pointTorqueDipoleMomentCoefficient$ is non-monotonic
in $H/(2a)$.

\section{Macroscopic suspension flow}
\label{Macroscopic suspension flow}

In this section we apply the results of the above analysis to determine the
average volume flux in a dilute suspension bounded by two parallel planar
walls.  The average flow can be driven by the macroscopic pressure gradient,
by lateral forces or by lateral torques applied to the particles.

To determine macroscopic equations governing suspension flow, we first
show that the far-field Hele--Shaw flow \refeq{Hele-Shaw flow} driven
by the dipolar pressure distribution \refeq{pressure dipole}
contributes to the macroscopic suspension velocity.  This behavior is
analogous to the 2D electrostatics, where the electric field produced
by induced dipoles contributes to the macroscopic electrostatic
displacement field (as discussed in Sec.\ \ref{subsection on
Macroscopic suspension flow}).

The electrostatic analogy can be directly applied to fluid flow
through a system of immobile particles in a channel.  In this case the
fluid transport is governed by the Darcy's equation
\begin{equation}
\label{Darcy's equation}
\averageSuspensionVelocity=-\permeability\macroscopicPressureGradient,
\end{equation}
which is a counterpart of the constitutive relation between the electric and
electrostatic-displacement fields.  In equation \refeq{Darcy's equation}
$\macroscopicPressureGradient$ denotes the macroscopic pressure gradient, and
$\averageSuspensionVelocity$ is the macroscopic velocity, defined as
fluid-flux density averaged across the channel.  The effective permeability
coefficient $\permeability$ plays a role similar to that of the effective
dielectric constant in the corresponding electrostatic system.  Equation
\refeq{Darcy's equation} can be obtained by combining Eq.\ \refeq{dipolar
expression for average suspension velocity} with the ``$\pressure\pressure$''
component of Eq.\ \refeq{generalized resistance relation}.

Note that relations \refeq{dipolar expression for average suspension
velocity} and \refeq{Darcy's equation} involve the macroscopic
pressure gradient (rather than the gradient of the external pressure
$\externalPressure$), to ensure that the constitutive relations
involve only local quantities and are independent of the boundary
conditions.

For a system of particles suspended in a fluid, suspension transport
is described by a constitutive equation that includes additional force
and torque contributions,
\begin{equation}
\label{macroscopic suspension flux}
H\averageSuspensionVelocity=
   \effectiveMobilityMatrixPT\bcdot\totForce
   +\effectiveMobilityMatrixPR\bcdot\totTorque
   -\effectiveMobilityMatrixPP\macroscopicPressureGradient,
\end{equation}
where $\effectiveMobilityMatrixPP$, $\effectiveMobilityMatrixPT$, and
$\effectiveMobilityMatrixPR$ are the effective mobility coefficients.

To derive Eq.\ \refeq{dipolar expression for average suspension
velocity} and determine the transport coefficients in Eqs.\
\refeq{Darcy's equation} and \refeq{macroscopic suspension flux}, we
consider the average pressure gradient and average velocity in a
periodic system representing a macroscopically uniform
quasi-two-dimensional medium.

\subsection{Periodic Green's function}

We start our analysis by deriving appropriate expressions for the
periodic Green's functions for Stokes flow in a parallel-wall channel.
In the present paper, our formulation is used as a theoretical tool to
obtain the relation between the quantities
$\averageParticleDipoleMoment$ and $\averageSuspensionVelocity$.  Our
explicit results are also applied in numerical calculations presented
in Sec.\ \ref{Transport coefficients}.  More generally, our formulas
for the periodic Green's functions can be employed in
Stokesian-dynamics and boundary-integral algorithms for parallel-wall
geometry.

\subsubsection{Near-field and far-field contributions}

In our approach, the periodic flow and pressure Green's functions
$\periodicGreenT$ and $\periodicGreenQ$ are evaluated by splitting them into
the asymptotic Hele--Shaw parts $\periodicGreenTasymptotic$ and
$\periodicGreenQasymptotic$, and quickly convergent lattice sums of
3D corrections to the asymptotic Hele--Shaw behavior,
\begin{subequations}
\label{periodic Green's functions - near and far-field contributions}
\begin{equation}
\label{periodic Green's function T - near and far-field contributions}
\periodicGreenT(\br,\br')
   =\periodicGreenTasymptotic(\br,\br')  
   +\sum_\bn\delta\GreenT(\br,\br'_\bn),
\end{equation}
\begin{equation}
\label{periodic Green's function Q - near and far-field contributions}
\periodicGreenQ(\br,\br')
   =\periodicGreenQasymptotic(\br,\br')
   +\sum_\bn\delta\GreenQ(\br,\br'_\bn).
\end{equation}
\end{subequations}
Here $\bn=(n_x,n_y)$ (with $n_x,n_y=0,\pm1\dots$) are the indices of
the periodic lattice, 
\begin{equation}
\label{lattice-site positions}
\br'_\bn=\br'+n_x L_x\ex+n_y L_y\ey
\end{equation}
are the positions of the periodic images of the source point $\br'$, and
$L_x$ and $L_y$ are the lattice constants.  The near-field
contributions $\delta\GreenT$ and $\delta\GreenQ$ are defined as the
differences between the exact and asymptotic non-periodic Green's
functions,
\begin{subequations}
\label{near field contributions}
\begin{equation}
\label{near field contribution T}
\delta\GreenT(\br,\br')=\GreenT(\br,\br')-\GreenTasymptotic(\br,\br'),
\end{equation}
\begin{equation}
\label{near field contribution Q}
\delta\GreenQ(\br,\br')=\GreenQ(\br,\br')-\GreenQasymptotic(\br,\br').
\end{equation}
\end{subequations}
Since the flow and pressure fields in a parallel-wall channel tend to
the asymptotic Hele--Shaw form exponentially on the lengthscale $H$
\cite{Liron-Mochon:1976,Bhattacharya-Blawzdziewicz-Wajnryb:2006}, the
near-field contributions \refeq{near field contributions}
exponentially vanish at large lateral distances $\delta\rho$ between
the field and source points $\br$ and $\br'$.  Our numerical tests
\cite{Bhattacharya-Blawzdziewicz-Wajnryb:2006} indicate that the exact
and asymptotic Green's functions are nearly identical for
$\delta\rho/H\gtrsim 3$.  Therefore, in practical calculations only a
small number of terms need to be evaluated to determine the lattice
sums in Eqs.\ \refeq{periodic Green's functions - near and far-field
contributions} with high accuracy.

We note that explicit expressions for all terms in Eqs.\ \refeq{near field
contributions} are known [cf.\ Eq.\ \refeq{non-periodic asymptotic Green's
functions} and the results in Appendix \ref{Expressions for velocity and
pressure Green's functions}].  Analytic formulas for the asymptotic Hele--Shaw
Green's functions $\periodicGreenTasymptotic$ and $\periodicGreenQasymptotic$
are derived in the following section and in Appendix \ref{Periodic formulation
for two-dimensional Laplace's equation}.

\subsubsection{Far-field Green's functions for periodic system}

To evaluate the far-field components $\periodicGreenTasymptotic$ and
$\periodicGreenQasymptotic$ of the periodic Green's functions
\refeq{periodic Green's functions - near and far-field contributions}
we start from the direct lattice sums
\begin{subequations}
\label{direct sums for asymptotic periodic Green's functions}
\begin{equation}
\label{direct sum for asymptotic periodic Green's function T}
\periodicGreenTasymptotic(\br,\br')=\sum_\bn\GreenTasymptotic(\br,\br'_\bn),
\end{equation}
\begin{equation}
\label{direct sum for asymptotic periodic Green's function Q}
\periodicGreenQasymptotic(\br,\br')=\sum_\bn\GreenQasymptotic(\br,\br'_\bn).
\end{equation}
\end{subequations}
Combining relations \refeq{direct sums for asymptotic periodic Green's
functions} with \refeq{non-periodic asymptotic Green's functions}
allows us to express the hydrodynamic periodic Green's functions in
the Hele--Shaw regime in terms of the periodic solution of the
corresponding electrostatic problem.   Accordingly, we have 
\begin{subequations}
\label{periodic asymptotic Green's functions}
\begin{equation}
\label{periodic asymptotic Green's function T}
\periodicGreenTasymptotic(\br,\br')
   =-\halff \eta^{-1}z(H-z)\bnabla\periodicGreenQasymptotic(\br,\br'),
\end{equation}
\begin{equation}
\label{periodic asymptotic Green's function Q}
\periodicGreenQasymptotic(\br,\br')
   =-3\pi^{-1}H^{-3}\bnabla\Wigner(\brho-\brho')z'(H-z'),
\end{equation}
\end{subequations}
where the Wigner potential $\Wigner(\brho-\brho')$, is the periodic
solution of the Poisson equation
\begin{equation}
\label{periodic Poisson equation}
\nabla^2\Wigner(\brho-\brho')=-2\pi\left[
    \sum_\bn\delta(\brho-\brho'_\bn)-\unitCellArea^{-1}
         \right].
\end{equation}
Here $\brho'_\bn$ is the lateral component of the lattice vector
\refeq{lattice-site positions}, and $\unitCellArea=L_x L_y$ is the
area of a unit cell \cite{convergence_note}.

The Wigner potential $\Wigner(\brho-\brho')$ can be determined using
standard Ewald summation techniques
\cite{Frenkel-Smit:2002,Cichocki-Felderhof:1989a}.  Well-developed
accelerated algorithms for calculating this function are also
available \cite{Frenkel-Smit:2002}.  Equations \refeq{periodic Green's
functions - near and far-field contributions} and \refeq{direct sums
for asymptotic periodic Green's functions} thus reduce the problem of
evaluating the doubly-periodic 3D hydrodynamic Green's
functions to a much simpler scalar problem \refeq{periodic Poisson
equation}.  Explicit expressions for the Wigner function and its
multipolar projections are given in Appendix \ref{Periodic formulation
for two-dimensional Laplace's equation}.  We also derive there Ewald
sums for the hydrodynamic Green's functions \refeq{periodic asymptotic
Green's functions}.

\subsection{Average flow produced by a point force}
\label{Expression for average flow}

The asymptotic periodic pressure Green's function \refeq{periodic
asymptotic Green's function Q} is proportional to the gradient of the
periodic Wigner function $\Wigner$.  Thus $\periodicGreenQasymptotic$
is normalized to yield zero average value over a unit cell (but any constant can be added without changing the system dynamics).

In contrast, the velocity Green's function $\periodicGreenT$ is fully
determined by Eqs.\ \refeq{periodic Green's function T - near and
far-field contributions} and \refeq{periodic asymptotic Green's
functions}, with no gauge constants involved. Adding a constant would
violate the boundary conditions on the channel walls---thus, the
average flow field in a wall-bounded periodic system cannot be set
independently.  Instead, the average flow is a function of the applied
pressure drop and the induced-force distribution (unlike the
corresponding behavior in the infinite space).

As a key step in our analysis of suspension flow in a channel, we
derive an expression for the average flow $\averagePointForceFlow$
produced by a lateral point force $\lateralPointForce$ applied at a
point $\br'=(\brho',\bz')$. (Only a lateral force needs to be
considered, because a transverse force $\bF_\perp=F_\perp\ez$ does not
produce an average flow, by symmetry).  The overall pressure drop in
the system is assumed to vanish.

The average flow is given by the integral over a unit cell
\begin{equation}
\label{average flow produced by point force - definition}
\averagePointForceFlow
   =\unitCell^{-1}\int_\unitCell\pointForceFlow(\br)\diff\br
\end{equation}
of the flow field
\begin{equation}
\label{flow field produced by point force}
\pointForceFlow(\br)=\periodicGreenT(\br,\br')\bcdot\lateralPointForce
\end{equation}
produced by the applied force.  To determine the volume integral in
Eq.\ \refeq{average flow produced by point force - definition} we use
the decomposition \refeq{periodic Green's function T - near and
far-field contributions} of the Green's function $\periodicGreenT$
into the near-field and far-field components.  The far-field
contribution vanishes because
\begin{equation}
\label{unit-cell integral of far-field Green's function}
\int_\unitCell\periodicGreenTasymptotic(\br,\br')=0,
\end{equation}
owing to Eq.\ \refeq{periodic asymptotic Green's function T} and
the periodicity of the pressure Green's function
$\periodicGreenQasymptotic$.  Therefore, the average flow
$\averagePointForceFlow$ is associated with the integral of the
near-field contribution
\begin{equation}
\label{near-field contribution to the point-force flow}
\delta\pointForceFlow(\br)=
   \sum_\bn\delta\GreenT(\br,\br_\bn')\bcdot\lateralPointForce
\end{equation}
over the unit cell $\unitCell$.  Using the invariance of
$\delta\GreenT(\br,\br')$ with respect to lateral translations, this
integral can be represented as
\begin{equation}
\label{integral of single near-field contribution over whole channel}
\averagePointForceFlow
   =\unitCell^{-1}\int_\unitCell\delta\pointForceFlow(\br)\diff\br
   =\unitCell^{-1}\int_\wholeSpace
                  \delta\GreenT(\br,\br')\bcdot\lateralPointForce,
\end{equation}
where $\wholeSpace=\sum_\bn\unitCell(\bn)$ is the whole infinite
volume of the channel.

The integral \refeq{integral of single near-field contribution over
whole channel} can be related to the dipolar strength \refeq{dipolar
moment pressure produced by point force} of the asymptotic pressure
distribution \refeq{far field pressure produced by point force} by
inserting under the integration sign the identity tensor
$\mathsfb{I}=\bnabla\br$ and integrating by parts.  Since the boundary
term vanishes, owing to the boundary conditions on the walls and the
rapid decay of $\delta\GreenT(\br,\br')$ for $\rho\to\infty$, we find
that
\begin{equation}
\label{near field integral as moment of divergence}
\averagePointForceFlow=
   -\unitCell^{-1}\int_\wholeSpace\br\bnabla\bcdot
    \delta\GreenT(\br,\br')\bcdot\lateralPointForce\diff\br.
\end{equation}
Noting that the exact Green's function in Eq.\ \refeq{near field
contribution T} is divergence-free and using relations
\refeq{non-periodic asymptotic Green's functions}, \refeq{far field
pressure produced by point force}, and \refeq{Poisson equation for
dipolar pressure} for the asymptotic contribution we obtain the
following fundamental result
\begin{equation}
\label{average point-force flow in terms of dipolar moment}
   \averagePointForceFlow=\channelPermeability\unitCell^{-1}H
       \pressureDipoleMoment,
\end{equation}
where $\pressureDipoleMoment$ is the dipole moment \refeq{dipolar
moment pressure produced by point force}, and $\channelPermeability$
is the permeability coefficient for the particle-free channel
\refeq{channel permeability}.  In an explicit form we have
\begin{equation}
\label{explicit expression for average point-force flow}
\averagePointForceFlow=\halff\eta^{-1}
    z'(H-z')\unitCell^{-1}\lateralPointForce.
\end{equation}
Relation \refeq{explicit expression for average point-force flow} is
consistent with the average fluid velocity produced by a planar force
distribution 
\begin{equation}
\label{planar force density}
f_\surface(\br)=\unitCell^{-1}H\delta(z-z')\lateralPointForce.
\end{equation}

\subsection{Average flow in the particle presence}

The macroscopic fields $\bnablaLat\averagePressure$ and
$\averageSuspensionVelocity$ that appear in the effective-medium equations
\refeq{Darcy's equation} and \refeq{macroscopic suspension flux} can be
identified with the volume averages of the microscopic pressure gradient
$\bnablaLat p$ and velocity $\bv$ over a unit cell $\unitCell$,
\begin{subequations}
\label{macroscopic pressure drop and velocity}
\begin{equation}
\label{macroscopic pressure drop}
\bnablaLat\averagePressure=\unitCell^{-1}\int_\unitCell \bnablaLat p\diff\br,
\end{equation}
\begin{equation}
\label{macroscopic velocity}
\averageSuspensionVelocity=\unitCell^{-1}\int_\unitCell \bv\diff\br.
\end{equation}
\end{subequations}
To determine the macroscopic transport coefficients in
equations \refeq{Darcy's equation} and \refeq{macroscopic suspension
flux}, we thus need to derive appropriate expressions for these
averages.

In this section we consider a system of $N$ particles at positions
$\bR_i$ ($i=1,\ldots,N$) in the unit cell.  The particles are
represented by the corresponding induced-force distributions
$\bF_i(\br)$.

It is convenient to represent the pressure and flow fields
\refeq{boundary integrals for velocity and pressure fields}
(generalized to a multiparticle system) as the superpositions of the
external and scattered contributions
\begin{subequations}
\label{external and scattered fields}
\begin{equation}
\label{external and scattered fields - pressure}
p=\externalPressure+\sum_{i=1}^N\scatteredPressurePart_i,
\end{equation}
\begin{equation}
\label{external and scattered fields - flow}
\bv=\externalVelocity+\sum_{i=1}^N\scatteredVelocityPart_i.
\end{equation}
\end{subequations}

We assume that the applied pressure gradient
$\bnabla\externalPressure$ is constant in space and has only lateral
components $x$ and $y$. The corresponding external velocity field
\refeq{external parabolic flow} depends only on the transverse
coordinate $z$.  The scattered flow and pressure fields
$\scatteredVelocityPart$ and $\scatteredPressurePart$ are periodic,
and they are given by the expressions
\begin{subequations}
\label{definition of perturbation pressure and velocity}
\begin{equation}
\label{definition of perturbation pressure}
\scatteredPressurePart_i=
   \int_\unitCell\periodicGreenQ(\br,\br')\bcdot\bF_i(\br')\diff\br',
\end{equation}
\begin{equation}
\label{definition of perturbation velocity}
\scatteredVelocityPart_i=
   \int_\unitCell\periodicGreenT(\br,\br')\bcdot\bF_i(\br')\diff\br'.
\end{equation}
\end{subequations}

By integrating the lateral gradient of the pressure \refeq{external
and scattered fields - pressure} over the unit cell $\unitCell$ we
find
\begin{equation}
\label{average pressure drop}
\bnablaLat\averagePressure=\bnabla\externalPressure,
\end{equation}
where the integrals of the perturbation-pressure terms
$\bnablaLat\scatteredPressurePart_i$ vanish by periodicity of
$\scatteredPressurePart$.  Equation \refeq{average pressure drop} is
important, because it allows us to express the macroscopic flow and
particle motion in a channel in terms of the macroscopic pressure
(rather than the external pressure) in the constitutive relations
derived in Sec.\ \ref{Transport coefficients}.

Integrating relation \refeq{external and scattered fields - flow} with
the external flow given by Eq.\ \refeq{external parabolic flow} yields
\begin{equation}
\label{average velocity}
\averageSuspensionVelocity=-\channelPermeability\bnabla\externalPressure
   +\unitCell^{-1}\sum_{i=1}^N\int_{\unitCell}
   \scatteredVelocityPart_i(\br)\diff\br.
\end{equation}
The integral on the right-hand-side of the above equation can be
evaluated using the result \refeq{average point-force flow in terms of
dipolar moment} for the average velocity produced by a point force.
Combining point-force results \refeq{average flow produced by point
force - definition} and \refeq{flow field produced by point force}
with \refeq{definition of perturbation velocity}, and taking into
account that the transverse force components do not contribute to the
average flow, we find
\begin{equation}
\label{average velocity produced by induced forces}
\averageSuspensionVelocity=\channelPermeability
    (-\bnabla\externalPressure
    +\unitCell^{-1}H\sum_{i=1}^N\dipoleMoment_i),
\end{equation}
where $\dipoleMoment_i$ is the dipole moment \refeq{expression for
dipolar moment of induced force distribution} of particle $i$.
Defining the average dipole moment
\begin{equation}
\label{average dipolar moment}
\averageParticleDipoleMoment=N^{-1}\sum_{i=1}^N\dipoleMoment_i,
\end{equation}
relation \refeq{dipolar expression for average suspension velocity}
for the average flow is thus obtained.

According to the above derivation, equation \refeq{dipolar expression
for average suspension velocity} is valid for arbitrary particle
densities, provided that the induced forces $\bF_i$ are evaluated with
the multiparticle hydrodynamic interactions properly taken into
account.  In the low-density limit, the average dipole moment
\refeq{average dipolar moment} can be expressed in terms of the dipole
moment of an isolated particle $\dipoleMoment(Z)$,
\begin{equation}
\label{low density average dipolar moment}
\suspensionDensityA\averageParticleDipoleMoment
   =\int_{a}^{H-a}\suspensionDensityV(Z)\dipoleMoment(Z)\diff Z,
\end{equation}
where $\suspensionDensityV(Z)$ is the local particle number density
per unit volume, averaged over the lateral position within a unit
cell.

\section{Transport coefficients}
\label{Transport coefficients}

In this section we use the relation for the average velocity
\refeq{dipolar expression for average suspension velocity} to obtain
the effective macroscopic equations for fluid and particle transport
in a parallel-wall channel.  In a dilute-suspension regime, the
macroscopic transport equations are obtained by combining
\refeq{dipolar expression for average suspension velocity} with the
friction relation \refeq{generalized resistance relation} or mobility
relation \refeq{generalized mobility relation}.  At higher particle
concentrations, we use \refeq{dipolar expression for average
suspension velocity} in combination with linear constitutive relations
between the dipole moment and the macroscopic forcing for a system of
interacting particles.

We consider here two important cases.  In Sec.\ \ref{Fixed bed of particles}
we discuss fluid transport through a fixed bed of particles, and in Sec.\
\ref{Suspension transport} we examine transport of a suspension of freely
moving particles.

\begin{figure}
\includegraphics*{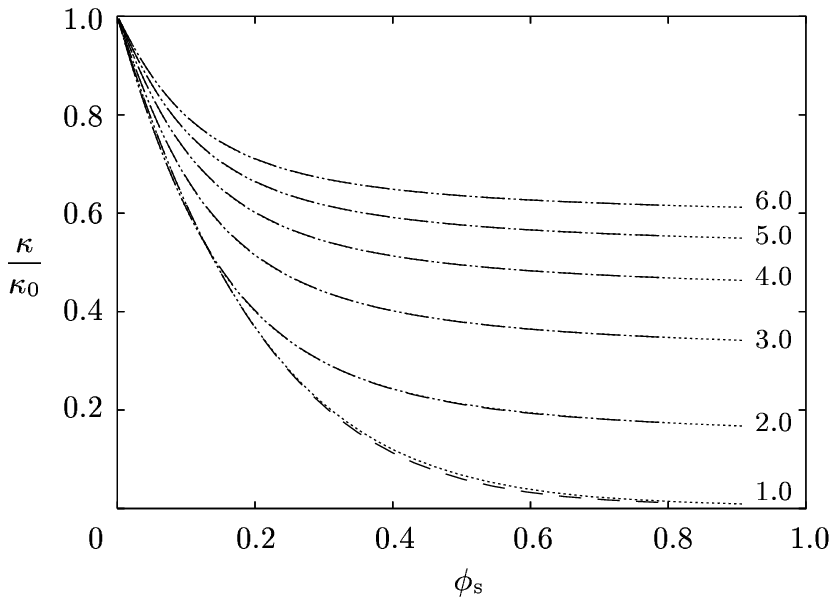}
\caption{\small{
Effective permeability \refeq{expression for permeability} of a
channel with a regular particle array adsorbed on the lower wall,
versus area fraction $\areaFraction$, for several values of normalized
channel width $H/(2a)$ (as labeled).  Hexagonal particle arrangement
(dotted lines); square arrangement (dashed lines).
}}
\label{permeability plot}
\end{figure}

\begin{figure}
\includegraphics*{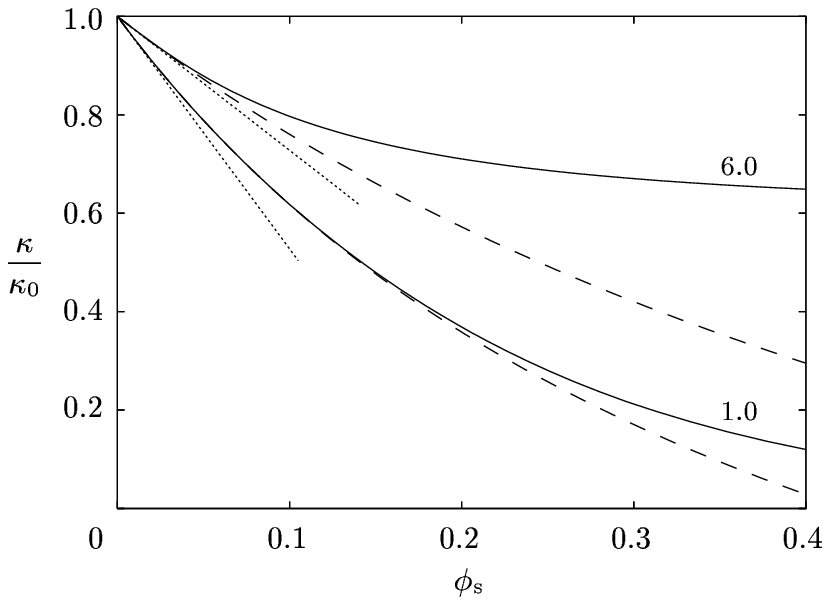}
\caption{\small{
Effective permeability \refeq{expression for permeability} of a
channel with a hexagonal particle array adsorbed on the lower wall,
versus area fraction $\areaFraction$, for two values of normalized
channel width $H/(2a)$ (as labeled).  Exact result (solid lines);
low-density limit \refeq{low-density behavior of permeability -
adsorbed particles} (dotted); Clausius--Mossotti approximation
\refeq{Clausius--Mossotti approximation} (dashed).
}}
\label{Clausus--Mossotti plot}
\end{figure}

\begin{figure}[t]
\includegraphics*{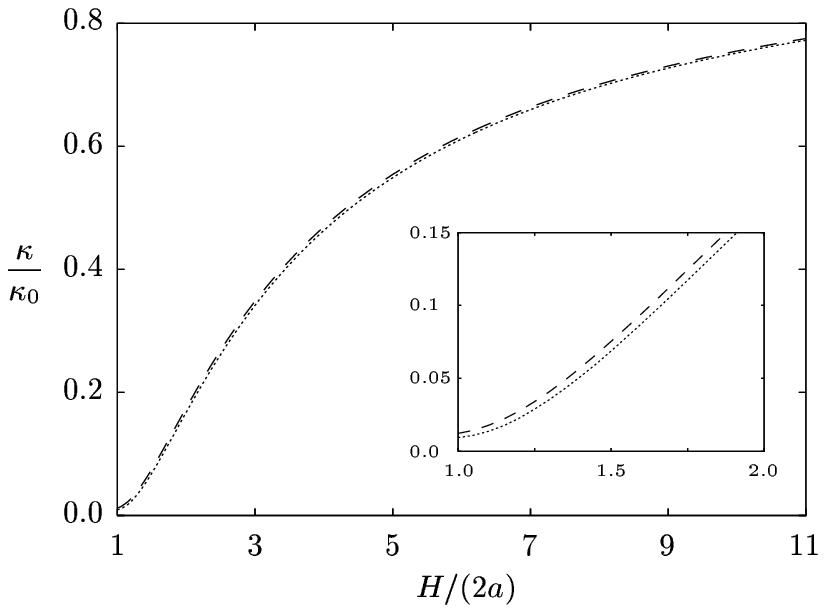}
\caption{\small{
Effective permeability \refeq{expression for permeability} of a
channel with a closed-packed regular particle array adsorbed on the
lower wall, versus normalized channel width $H/(2a)$.  Hexagonal
particle arrangement (dotted lines); square arrangement (dashed lines).
Inset shows a blowup of the plot for small values of $H/(2a)$.
}}
\label{contact-value plot}
\end{figure}

\begin{figure}[t]
\includegraphics*{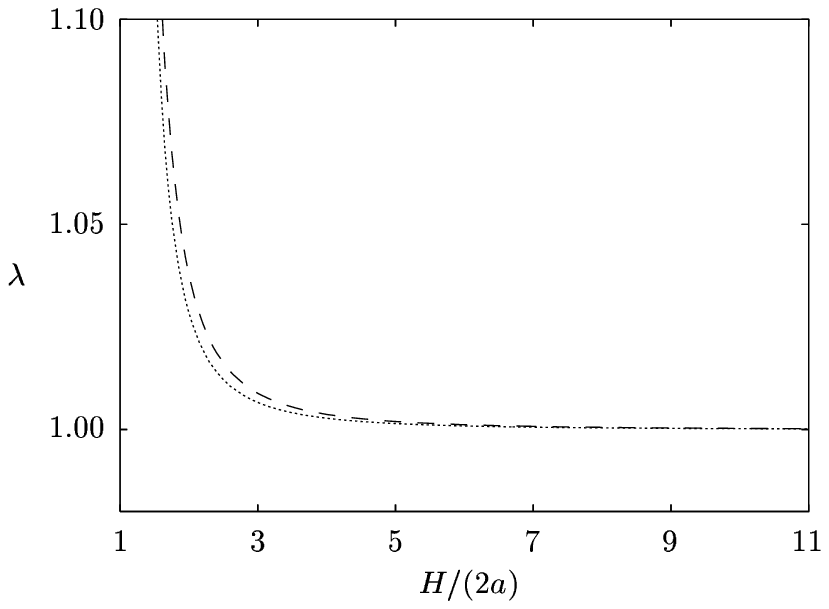}
\caption{\small{
Fluid-flux ratio \refeq{fluid-flux ratio} for the same two systems as
those represented in Fig.\ \ref{contact-value plot}.
}}
\label{contact-value scaled plot}
\end{figure}

\subsection{Fixed bed of particles}
\label{Fixed bed of particles}

\subsubsection{Permeability coefficient}
\label{subsection permeability coefficient}

We begin our analysis of fluid transport through a fixed particle
array by considering a low-density system.  In the low-density domain,
the dipole moment $\averageParticleDipoleMoment$ contribution to the
average flow \refeq{dipolar expression for average suspension
velocity} can be obtained by averaging the polarizability component of
the generalized resistance relation \refeq{generalized resistance
relation} over the particle distribution.  Taking into account that
$\particleVelocity=\particleAngularVelocity=0$ we find that
\begin{equation}
\label{friction pressure-pressure for periodic system}
\smallfrac{1}{12}H^3\averageParticleDipoleMoment
   =\averageResistanceMatrixPP\bcdot\eta^{-1}\macroscopicPressureGradient,
\end{equation}
where 
\begin{equation}
\label{average one-particle polarization}
\averageResistanceMatrixPP=
  \suspensionDensityA^{-1}
  \int_a^{H-a}\suspensionDensityV(Z)\resistanceMatrixPP(Z)\diff Z
\end{equation}
is the mean value of the $\resistanceMatrixPP$ component of the
generalized resistance matrix in Eq.\ \refeq{generalized resistance
relation}.  The driving force in the polarizability relation
\refeq{friction pressure-pressure for periodic system} is the gradient
of the macroscopic pressure $\macroscopicPressureGradient$, which at
low densities is identical to $\bnablaLat\externalPressure$.  By
inserting \refeq{friction pressure-pressure for periodic system} into
\refeq{dipolar expression for average suspension velocity} we obtain
Darcy's equation \refeq{Darcy's equation} with the permeability
coefficient of the form
\begin{equation}
\label{expression for permeability}
\permeability
  =\channelPermeability(1-
   12\suspensionDensityA H^{-3}\eta^{-1}\averageResistanceMatrixPP),
\end{equation}
where definition \refeq{channel permeability} was used to factor out
the permeability of particle-free channel $\channelPermeability$.

The polarizability relation \refeq{friction pressure-pressure for
periodic system} and expression \refeq{expression for permeability}
for the channel permeability are valid not only for
dilute-suspensions, but also for arbitrary particle concentrations,
provided that the average \refeq{average one-particle polarization} is
replaced by the corresponding relation applicable in the
high-concentration regime.  As in other problems of field propagation
through random media, the polarizability coefficient
$\averageResistanceMatrixPP$ can be expressed in terms of cluster
integrals that involve pair, triplet, and higher-order resistance
functions.  Derivation of such a relation requires applying an
appropriate renormalization procedure in which the external pressure
is replaced by the macroscopic pressure in order to obtain absolutely
convergent results for $\averageResistanceMatrixPP$.  Alternatively,
we can get the polarizability from the dipole moment of particles in a
channel with periodic boundary conditions in the lateral
directions. Numerical results described in the following section have
been obtained using the latter technique.

\subsubsection{Numerical results}
\label{Numerical results - permeability}

To illustrate the effect of immobile particles on the permeability of
a parallel-wall channel we present results for a particle monolayer
adsorbed on one of the walls.  Figures \ref{permeability
plot}--\ref{contact-value scaled plot} show the permeability
coefficient (normalized by the permeability of a particle-free
channel) for hexagonal and square particle arrays.  For the square and
hexagonal symmetry, the tensorial permeability coefficient is
isotropic
\begin{equation}
\label{isotropic permeability}
\permeability=\scalarPermeability\lateralUnitTensor,
\end{equation}
where $\lateralUnitTensor$ is the lateral unit tensor \refeq{lateral
unit tensor}.

Figure \ref{permeability plot} illustrates the dependence of the
permeability coefficient $\scalarPermeability$ on the particle area
fraction $\areaFraction=\suspensionDensityA\pi a^2$ for several values
of channel width.  The results indicate that the permeabilities of the
hexagonal and square arrays are nearly identical functions of
$\areaFraction$---only for a channel with $H/(2a)\approx 1$ there is a
noticeable difference, especially at high surface coverage
$\areaFraction$.

In Fig.\ \ref{Clausus--Mossotti plot} our numerical results for
hexagonal arrays are compared to the low-density limiting behavior
\begin{equation}
\label{low-density behavior of permeability - adsorbed particles}
\frac{\scalarPermeability}{\channelPermeability}
   =1-\normalizedResistanceCoefficientPP\areaFraction,
\end{equation}
where the normalized one-particle polarizability coefficient
$\normalizedResistanceCoefficientPP$ is given by Eq.\
\refeq{Normalized one-particle resistance coefficient pp}.  Relation
\refeq{low-density behavior of permeability - adsorbed particles}
follows from Eqs.\ \refeq{average one-particle polarization} and
\refeq{expression for permeability} applied to a particle monolayer.
For particles positioned in the midplane of the channel the particle
density, averaged over the lateral directions, is
$n(Z)=\suspensionDensityA\delta(Z-\halff H)$, where
$\suspensionDensityA=N/(L_x L_y)$, and $N$ denotes the number of
particles in a unit cell.  For more general particle distributions,
the average value of $\normalizedResistanceCoefficientPP$ would appear
in \refeq{low-density behavior of permeability - adsorbed particles}.

In Fig.\ \ref{Clausus--Mossotti plot} we also plot the
Clausius--Mossotti approximation
\begin{equation}
\label{Clausius--Mossotti approximation}
\frac{\scalarPermeability}{\channelPermeability}
  =\frac{1-\halff\normalizedResistanceCoefficientPP\areaFraction}
        {1+\halff\normalizedResistanceCoefficientPP\areaFraction},
\end{equation}
which is a generalization of the classical electrostatic
Clausius--Mossotti formula \cite{Jackson:1999} to our present problem.
The results indicate that for tightly confined systems with
$H/(2a)\approx1$, the approximation \refeq{Clausius--Mossotti
approximation} is quite accurate in the area-fraction range
$\areaFraction\lesssim0.2$.  For weaker confinements, the range of
validity of the Clausius--Mossotti formula is smaller.

The permeability of close-packed arrays (with the close-packing area
fraction $\areaFraction=\sqrt{3}\pi/6$ for hexagonal and
$\areaFraction=\pi/4$ for square ordering) is plotted in Fig.\
\ref{contact-value plot} versus the dimensionless wall separation
$H/(2a)$.  The results indicate that for $H/(2a)\approx1$, the
permeability coefficient is reduced to about 1\,\% of the permeability
$\channelPermeability$ of a particle-free channel with the same width.
Such a significant reduction of the channel permeability was observed
in recent experiments
\cite{Sung-Vanapalli-Mukhija-McKay-$Mirecki_Millunchick$-Burns-Solomon:2008}.

At weaker confinements, the reduction of the permeability is much
smaller, but it is still quite significant for $H/(2a)\lesssim10$.
For moderate and large channel widths the hindrance of fluid flow can
be accurately accounted for in terms of the reduced effective channel
width.  In this approximation, the particle array is replaced by an
equivalent solid slab of width $\zNoSlip$, occupying the region $0\le
z\le\zNoSlip$.  The permeability of the narrowed channel is
\begin{equation}
\label{permeability of channel of effective width}
\effectiveWidthPermeability=\smallfrac{1}{12}\eta^{-1}H_\effective^2\,,
\end{equation}
where
\begin{equation}
\label{effective channel width}
\Heffective=H-\zNoSlip
\end{equation}
is the effective channel width.  

The accuracy of the effective-width approximation \refeq{permeability
of channel of effective width} can be estimated from a plot of the
ratio
\begin{equation}
\label{fluid-flux ratio}
\fluidFluxRatio=\frac{H\scalarPermeability}
                {H_\effective\,\effectiveWidthPermeability}
\end{equation}
of the fluid flux through the channel with adsorbed particles to the
fluid flux through a particle-free channel of the reduced width.  Such
a plot is shown in Fig.\ \ref{contact-value scaled plot} for closely
packed hexagonal and square arrays.  From our data we find
$\zNoSlip=0.907$ for the hexagonal particle array and $\zNoSlip=0.895$
for the square array.  With the above values we get $\lambda\approx1$
for $H/(2a)\gtrsim2$.  In this range of channel widths, the
approximation that neglects the roughness of a densely packed particle
layer adsorbed on the wall is thus accurate.  We note that our
findings are consistent with earlier investigations of fluid flow near
rough surfaces
\cite{%
Lecoq-Anthore-Cichocki-Szymczak-Feuillebois:2004,
Vinogradova-Yakubov:2006%
}.

\subsection{Suspension of freely moving particles}
\label{Suspension transport}

\subsubsection{Macroscopic constitutive equation}

The linear constitutive equation relating the macroscopic fluxes to
macroscopic forces in the dilute-suspension regime is obtained by
combining the dipolar expression for the average suspension velocity
\refeq{dipolar expression for average suspension velocity} with the
generalized mobility relation \refeq{generalized mobility relation}.
Relation \refeq{generalized mobility relation} is first averaged over
the particle distribution, which yields
\begin{equation}
\label{macroscopic polarizability equation}
\left[
   \begin{array}{c}
      \averageParticleVelocity\\
      \averageParticleAngularVelocity\\
      \smallfrac{1}{12}\eta^{-1}H^3 \averageParticleDipoleMoment
   \end{array}
\right]
=
\left[
   \begin{array}{ccr}
      \averageMobilityMatrixTT&\averageMobilityMatrixTR&
                               \averageMobilityMatrixTP\\
      \averageMobilityMatrixRT&\averageMobilityMatrixRR&
                               \averageMobilityMatrixRP\\
      \averageMobilityMatrixPT&\averageMobilityMatrixPR&
                              -\averageMobilityMatrixPP\\
   \end{array}
\right]
   \bcdot
\left[
   \begin{array}{c}
      \totForce\\
      \totTorque\\
      -\macroscopicPressureGradient
   \end{array}
\right],
\end{equation}
where
\begin{equation}
\label{average one-particle polarization - mobility}
\suspensionDensityA\averageMobilityMatrix^{AB}_\lowDensity=
  \int_a^{H-a}\suspensionDensityV(Z)\mobilityMatrix^{AB}(Z)\diff Z,
\end{equation}
similar to Eq.\ \refeq{average one-particle polarization} in the
friction-representation case.  In Eq.\ \refeq{macroscopic
polarizability equation} it is assumed that the same force $\totForce$
and torque $\totTorque$ act on all the particles in the system.  We
also assume that both $\totForce$ and $\totTorque$ have only the
lateral components.  As it has been done for immobile particles, in
Eq. \refeq{macroscopic polarizability equation} the external-pressure
gradient is replaced with the gradient of the macroscopic pressure
$\macroscopicPressureGradient$ to obtain a local constitutive relation
that can be generalized to dense systems.

By combining \refeq{dipolar expression for average
suspension velocity} and \refeq{macroscopic polarizability equation}
we find
\begin{equation}
\label{macroscopic constitutive equation for suspension}
\left[
   \begin{array}{c}
      \suspensionDensityA\averageParticleVelocity\\
      \suspensionDensityA\averageParticleAngularVelocity\\
      H\averageSuspensionVelocity
   \end{array}
\right]
=
\left[
   \begin{array}{ccr}
      \effectiveMobilityMatrixTT&\effectiveMobilityMatrixTR&
                               \effectiveMobilityMatrixTP\\
      \effectiveMobilityMatrixRT&\effectiveMobilityMatrixRR&
                               \effectiveMobilityMatrixRP\\
      \effectiveMobilityMatrixPT&\effectiveMobilityMatrixPR&
                               \effectiveMobilityMatrixPP\\
   \end{array}
\right]
   \bcdot
\left[
   \begin{array}{c}
      \totForce\\
      \totTorque\\
      -\macroscopicPressureGradient
   \end{array}
\right],
\end{equation}
where
\begin{subequations}
\label{effective mobilities}
\begin{equation}
\label{effective mobility AB not PP}
\effectiveMobilityMatrix^{AB}
  =\suspensionDensityA\averageMobilityMatrix^{AB},
\qquad 
  AB\not=\parabolic\parabolic,
\end{equation}
and
\begin{equation}
\label{effective mobility PP}
\effectiveMobilityMatrixPP
  =H\channelPermeability(1
    -12\suspensionDensityA H^{-3}\eta\averageMobilityMatrixPP).
\end{equation}
\end{subequations}
In the constitutive relation \refeq{macroscopic constitutive equation
for suspension}, the macroscopic fluxes are: the particle flux
$\suspensionDensityA\averageParticleVelocity$, particle angular flux
$\suspensionDensityA\averageParticleAngularVelocity$, and the
suspension volume flux $H\averageSuspensionVelocity$.

Similar to the immobile-particle case, relations \refeq{macroscopic
polarizability equation}, \refeq{macroscopic constitutive equation for
suspension}, and \refeq{effective mobilities} are valid at arbitrary
densities, provided that the low-density formula \refeq{average
one-particle polarization - mobility} for the generalized mobility
coefficients $\averageMobilityMatrix^{AB}$ is replaced with the
corresponding expression that is appropriate at high
densities. 

\subsubsection{Onsager reciprocal relations for the generalized 
mobility matrix $\effectiveMobilityMatrix$}
\label{Onsager reciprocal relations for the generalized mobility matrix}

In Sec.\ \ref{Polarizability coefficients} we have shown that at low particle
concentrations the generalized mobility/polarizability matrix
$\mobilityMatrix$ in Eq.\ \refeq{generalized mobility relation} satisfies the
Lorentz symmetry \refeq{Lorentz symmetry of generalized mobility matrix}.  The
same symmetry also applies to the average defined by Eq.\ \refeq{average
one-particle polarization - mobility}. It follows that the matrix of the
kinetic coefficients \refeq{effective mobilities} in the constitutive relation
\refeq{macroscopic constitutive equation for suspension} satisfies the Onsager
reciprocal relation
\begin{equation}
\label{Onsager symmetry}
\effectiveMobilityMatrix^{AB}=\effectiveMobilityMatrix^{BA\,\dagger}.
\end{equation}

While our derivation is given here only for dilute suspensions, one
can show that the symmetry relation \refeq{Onsager symmetry} is valid
at arbitrary concentrations.  The symmetry of the matrix
$\effectiveMobilityMatrix$ can be demonstrated using arguments similar
to the ones given in Sec.\ \ref{Polarizability coefficients}, but
applied to a periodic multiparticle system.

\begin{figure}
\includegraphics*{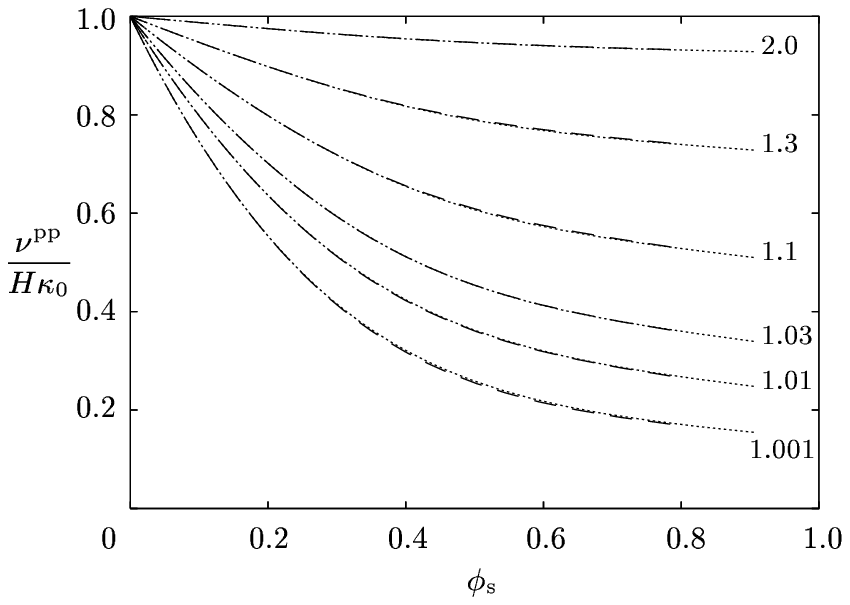}
\caption{\small{
Effective permeability \refeq{effective mobility PP} of a channel with
a regular particle array suspended in the midplane $Z=H/2$, versus
area fraction $\areaFraction$, for several values of normalized
channel width $H/(2a)$ (as labeled).  Hexagonal particle arrangement
(dotted lines); square arrangement (dashed lines).
}}
\label{free-permeability plot}
\end{figure}

\begin{figure}
\includegraphics*{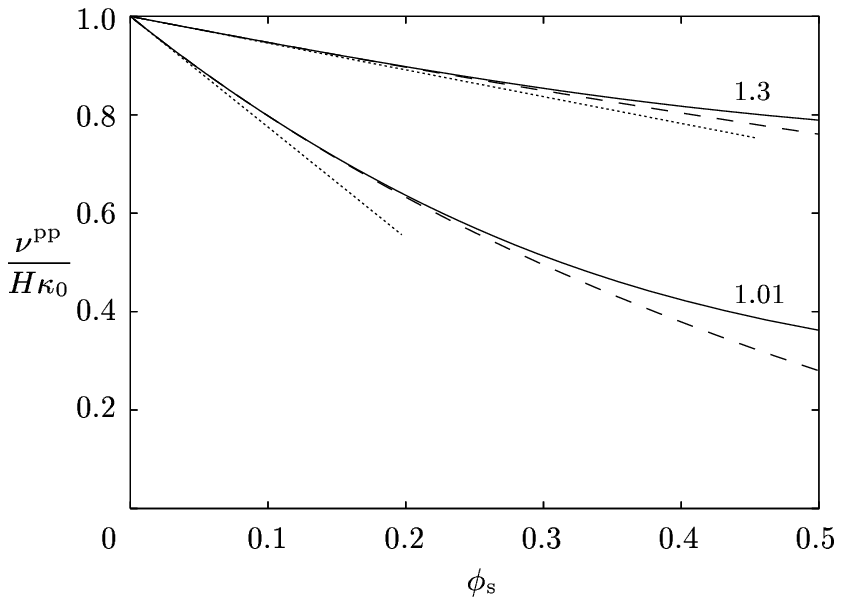}
\caption{\small{
Effective permeability \refeq{effective mobility PP} of a channel with
a hexagonal particle array freely suspended in the midplane $Z=H/2$,
versus area fraction $\areaFraction$, for two values of normalized
channel width $H/(2a)$ (as labeled).  Exact result (solid lines);
low-density limit \refeq{low-density behavior of permeability -
suspended particles} (dotted); Clausius--Mossotti approximation
\refeq{Clausius--Mossotti approximation free} (dashed).
}}
\label{free-Clausus--Mossotti plot}
\end{figure}

\begin{figure}
\includegraphics*{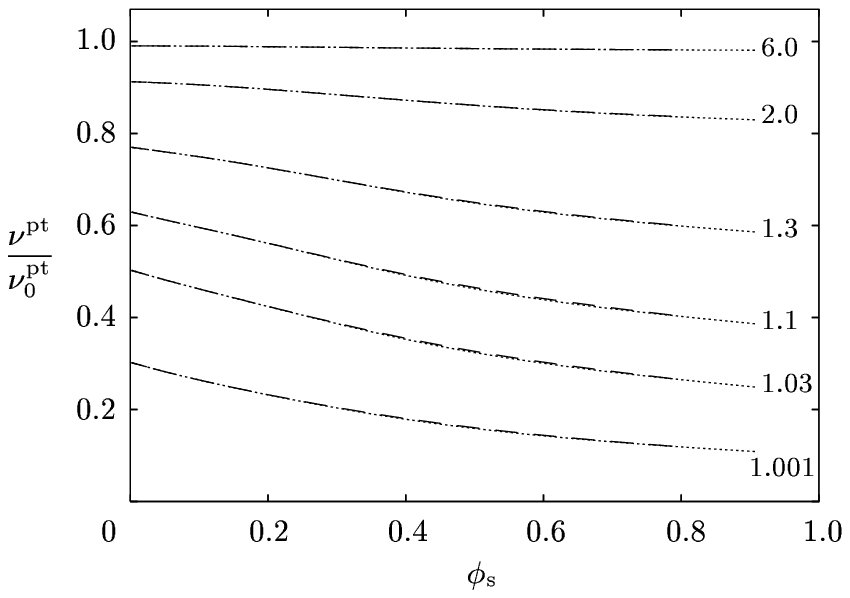}
\caption{\small
Effective mobility coefficient $\effectiveMobilityCoefficientPT$ for
an infinite particle array in the midplane $Z=H/2$, normalized by the
corresponding result for an array of point forces \refeq{flux
produced by point forces}, versus area fraction $\areaFraction$, for
several values of normalized channel width $H/(2a)$ (as labeled).
Hexagonal particle arrangement (dotted lines); square arrangement
(dashed lines).  The coefficient $\effectiveMobilityCoefficientPT$
describes the average suspension flow produced by a lateral force
applied to the particle array.
}
\label{nu pt multiparticle plot}
\end{figure}

\subsubsection{Numerical results}

At low suspension concentrations and, more generally, for suspensions that are
isotropic in the lateral directions, the transport coefficients
$\effectiveMobilityMatrix^{AB}$ are proportional to 2D isotropic
tensors
\begin{subequations}
\label{isotropic tensors and pseudo-tensors}
\begin{equation}
\label{isotropic tensors}
\effectiveMobilityMatrixPT=\effectiveMobilityCoefficientPT\lateralUnitTensor,
\qquad
\effectiveMobilityMatrixPP=\effectiveMobilityCoefficientPP\lateralUnitTensor,
\end{equation}
and
\begin{equation}
\label{isotropic pseudo-tensors}
\effectiveMobilityMatrixPR
  =\effectiveMobilityCoefficientPR\lateralAlternatingTensor,
\end{equation}
\end{subequations}
where $\lateralAlternatingTensor$ is the lateral alternating tensor
\refeq{lateral alternating tensor}.  Relations \refeq{isotropic
tensors and pseudo-tensors} are also satisfied for  hexagonal and
square particle lattices.

The effective channel permeability coefficient
$\effectiveMobilityCoefficientPP$, normalized by the permeability of a
particle-free channel, is plotted in Fig.\ \ref{free-permeability
plot} for infinite hexagonal and square particle arrays moving in the
midplane of the channel.  As for arrays of immobile particles, we find
that the permeabilities of the square and hexagonal arrays with the
same area fraction are nearly the same.

A comparison of the results depicted in Figs. \ref{permeability plot}
and \ref{free-permeability plot} indicates that the permeability of a
channel with particles freely suspended in the midplane $Z=H/2$ is
much higher than the corresponding permeability of a channel where the
particles are adsorbed at a wall.  According to Fig.\
\ref{free-permeability plot} the correction to the permeability due to
the particle presence is below 10\% in the regime $H/(2a)\gtrsim2$,
even for close-packed arrays.  This result is consistent with the
small value of the single-particle polarizability for a particle at
the symmetry plane $Z=H/2$, as illustrated in Fig.\ \ref{mu pp one
particle plot}.  Both the small particle polarizability and the minor
correction to the channel permeability stem from the small velocity
gradient in the midplane of the channel, which implies that the
suspended particles do not significantly perturb the fluid flow.

In Fig.\ \ref{free-Clausus--Mossotti plot} the results of the numerical
calculations for periodic particle arrays are compared with the low-density
expansion
\begin{equation}
\label{low-density behavior of permeability - suspended particles}
\frac{\effectiveMobilityCoefficientPP}{H\channelPermeability}
   =1-\normalizedMobilityCoefficientPP\areaFraction,
\end{equation}
where $\normalizedMobilityCoefficientPP$ is given by \refeq{Normalized
one-particle mobility coefficient pp}.  We also plot the
Clausius--Mossotti approximation
\begin{equation}
\label{Clausius--Mossotti approximation free}
\frac{\effectiveMobilityCoefficientPP}{H\channelPermeability}
  =\frac{1-\halff\normalizedMobilityCoefficientPP\areaFraction}
        {1+\halff\normalizedMobilityCoefficientPP\areaFraction}.
\end{equation}
A comparison of the results shown in Figs.\ \ref{Clausus--Mossotti
plot} and \ref{free-Clausus--Mossotti plot} indicate that the
Clausius--Mossotti approximation has a broader range of validity for
freely suspended particles than for the adsorbed ones.

The transport coefficients $\effectiveMobilityCoefficientPT$ and
$\effectiveMobilityCoefficientPR$, representing the average flow
produced by a lateral force and torque acting on the particles, are
plotted in Figs.\ \ref{nu pt multiparticle plot} and \ref{nu pr
multiparticle plot}.  The transport coefficients are shown normalized
by the respective results 
\begin{equation}
\label{flux produced by point forces}
\pointForceEffectiveMobilityCoefficient
   =\suspensionDensityA\pointForceDipoleMomentCoefficient(Z),
\end{equation}
\begin{equation}
\label{flux produced by point torque}
\pointTorqueEffectiveMobilityCoefficient
   =\suspensionDensityA\pointTorqueDipoleMomentCoefficient(Z)
\end{equation}
for the average flow produced by arrays of point forces and torques
applied to the suspending fluid at the particle positions $Z$, where
$\pointForceDipoleMomentCoefficient$ and
$\pointTorqueDipoleMomentCoefficient$ are given by relations
\refeq{point-force dipole-moment coefficient} and \refeq{point-torque
dipole-moment coefficient}.  The results are presented for square and
hexagonal particle arrays.  For the system driven by the lateral
force, the particles are in the midplane of the channel $Z=H/2$.  For
the lateral torque the rotational mobility coefficient
$\effectiveMobilityCoefficientPR$ is shown for $Z=a+\third(H-2a)$,
because the torque applied to the particles produces average flow only
for off-center positions.

\begin{figure}
\includegraphics*{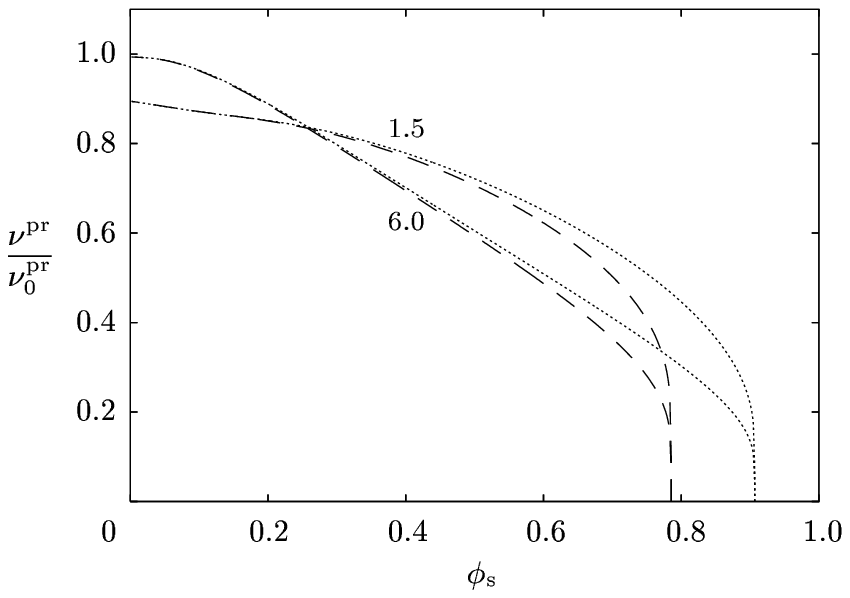}
\caption{\small
Effective mobility coefficient $\effectiveMobilityCoefficientPR$ for
an infinite particle array in the plane $Z=a+\third(H-2a)$, normalized
by the corresponding results for an array of point torques \refeq{flux
produced by point torque}, versus area fraction $\areaFraction$, for
several values of normalized channel width $H/(2a)$ (as labeled).
Hexagonal particle arrangement (dotted lines); square arrangement
(dashed lines).  The coefficient $\effectiveMobilityCoefficientPR$
describes the average suspension flow produced by a lateral torque
applied to the particle array.
}
\label{nu pr multiparticle plot}
\end{figure}

The results in Fig.\ \ref{nu pt multiparticle plot} indicate that for
$H/(2a)\gtrsim2$ the average flow produced in a channel by an external
force applied to the particles is well represented by the point-force
approximation \refeq{flux produced by point forces}, even for dense
particle arrays.  The normalized average flow in this regime is
insensitive both to the dimensionless channel width $H/(2a)$ and to
the area fraction $\areaFraction$.  For smaller values of $H/(2a)$,
the normalized average flow is smaller than the flow produced by point
forces.  However, a significant deviation of the normalized transport
coefficient
$\effectiveMobilityCoefficientPT/\pointForceEffectiveMobilityCoefficient$
from unity is observed only for particles with the diameter nearly
equal to the channel width.  This behavior is similar to the one seen
in Fig.\ \ref{free-permeability plot} for arrays driven by an external
pressure gradient. The low-density values are consistent with the
one-particle results represented in Fig.\ \ref{mu pt one particle
plot}.  The results for hexagonal and square lattices for a given area
fraction are nearly indistinguishable.

As depicted in Fig.\ \ref{nu pr multiparticle plot}, the average flow
produced by a torque $\totTorque$ applied to the particles strongly
depends on the particle area fraction, even for large wall
separations.  This is because particle rotation involves relative
motion of the surfaces of the spheres, whereas the motion of
force-driven particle monolayers does not involve any relative
particle displacements.  Moreover, we find that the results for the
square and hexagonal particle arrays are significantly different for
$\areaFraction\gtrsim 0.5$.  The transport coefficient
$\effectiveMobilityCoefficientPR$ vanishes at the close-packing area
fraction for a given system geometry, because the particle rotation is
arrested by the lubrication forces between the touching particles.  In
the low-density regime the results are independent of the particle
lattice.  The low-density values of the transport coefficients are
equivalent to the one-particle results represented in Fig.\ \ref{mu pr
one particle plot}.

\section{Conclusions}
\label{Conclusions}

We have presented a detailed analysis of the far-field scattered flow
produced by spherical particles in Stokes flow bounded by two parallel
planar walls.  Furthermore, we have examined the effect of the
far-field particle response to external forcing on the macroscopic
suspension flow.  Both the permeability of a system of fixed particles
in a channel and the macroscopic dynamics of a suspension of freely
moving particles were analyzed.  (We note that related ideas
were also explored in \cite{Bhattacharya:2008}, in the context of
molecular-dynamics simulations of the motion of nano-particles in a
confined fluid.%
)

For a system of fixed particles, the macroscopic fluid flux
is related to the macroscopic pressure via linear Darcy's law.  We have shown
that for a given macroscopic pressure gradient, the difference between the
fluid flux in the particle presence and in a particle-free channel can be
expressed in terms of the effective 2D dipole moment characterizing the
amplitude of the far-field Hele--Shaw dipolar scattered flow produced by the
particles.  From this amplitude we have evaluated the particle contribution to
the effective permeability coefficient.

A similar physical picture also applies to the flow of a suspension in
a channel.  However, in this case the macroscopic volume flux can be
produced not only by the macroscopic pressure gradient but also by the
external force or torque applied to the particles.  There are also
particle fluxes corresponding to the linear and angular particle
velocities.  Therefore, the constitutive relation for suspension flow
through a channel involves three forcing and three flux components.
The macroscopic fluxes and driving force are related through a
$3\times3$ matrix of (generally tensorial) transport coefficients.  We
have demonstrated that, with a proper normalization, this matrix is
symmetric, i.e., the transport coefficients satisfy the Onsager
reciprocal relations.

Our theoretical analysis has been supplemented by numerical results
for transport coefficients describing dynamics of square and hexagonal
particle arrays (for particle monolayers adsorbed on a wall and
monolayers of freely suspended particles).  We have shown that dense
arrays of tightly confined particles can reduce fluid flow through a
channel by as much as 99\%.

We have also proposed generalized Clausius--Mossotti formulas for the
channel permeability, both for fixed and freely suspended particles.
These formulas are analogous to the well-known electrostatic
Clausius--Mossotti approximation.  At moderate particle
concentrations, our expressions agree well with the numerical
calculations, especially for tightly confined particle arrays.

Our numerical results for the transport coefficients characterizing
macroscopic motion of regular particle arrays can be used to describe the
macroscopic deformation of finite-size regular 2D particle clusters.  As we
have shown in our recent paper \cite{Baron-Blawzdziewicz-Wajnryb:2008}, such
arrays exhibit a complex nonlinear dynamics that involves rearrangements of a
deformed particle lattice. There are also order-disorder transitions resulting
from lattice instabilities.  In future publications we will analyze these
problems using the macroscopic theory developed in this paper.  We will also
determine particle and fluid transport in suspensions of randomly distributed
particles.

Our analysis of suspension flow in parallel-wall channels with
periodic boundary conditions has also provided another important
result: we have derived explicit Ewald-summation formulas for the flow
and pressure periodic Green's functions in the parallel-wall geometry.
These formulas can be applied in Stokesian-dynamics and
boundary-integral simulations of suspension and emulsion flows in
narrow channels and slit pores, and in studies of the dynamics of
confined macromolecules.  

In future publications we will discuss the effect of the far-field flow on the
dynamics of macromolecules (e.g., polymer chains).  Earlier studies have
suggested that in a confined system the far-field contribution to the
hydrodynamic interactions between the chain segments can be neglected,
because, on average, this contribution vanishes by symmetry
\cite{Tlusty:2006,Balducci-Mao-Han-Doyle:2006}.
Our preliminary investigation based on the present results indicates that the
role of the far-field flow is much more subtle, although the far-field
hydrodynamic interactions do not change the Rouse scaling exponent for the
longest relaxation time (only the prefactor is affected).

\acknowledgments{We would like to acknowledge numerous useful discussions with
S.\ Bhattacharya at the early stages of this project.  We also acknowledge his
contribution to the derivation of the expressions for periodic Green's
functions, presented in Appendix \ref{Periodic formulation for two-dimensional
Laplace's equation}.  This work was supported by NSF CAREER grant CTS-0348175;
EW was also supported by Polish Ministry of Science grant N501 020 32/1994.}

\appendix

\section{Velocity and pressure Green's functions $\GreenT$ and $\GreenQ$}
\label{Expressions for velocity and pressure Green's functions}

In this Appendix we present our explicit expressions for the velocity
and pressure Green's functions $\GreenT$ and $\GreenQ$ for Stokes flow
between two parallel planar walls.  The expressions are obtained using
our Cartesian-representation approach introduced in
\cite{%
Bhattacharya-Blawzdziewicz-Wajnryb:2005,
Bhattacharya-Blawzdziewicz-Wajnryb:2005a%
}.
In our previous papers explicit results were given only for multipolar
projections of the Green's tensor $\GreenT$.  

The Green's functions for Stokes flow between two parallel walls can be
expressed as a sum of the free-space part and the wall contribution,
\begin{subequations}
\label{free-space and wall contributions to Green functions}
\begin{equation}
\label{free-space and wall contributins to velocity Green function}
\GreenT(\br,\br')=\OseenTensor(\br-\br')+\wallGreenT(\br,\br')
\end{equation}
\begin{equation}
\label{free-space and wall contributins to pressure Green function}
\GreenQ(\br,\br')=\OseenQ(\br-\br')+\wallGreenQ(\br,\br')
\end{equation}
\end{subequations}
where
\begin{equation}
\label{Oseen Green functions}
\OseenTensor(\br)=\frac{1}{8\pi\eta r}(\identityTensor+\br\br),
\quad
\OseenQ(\br)=\frac{1}{8\pi\eta r}
\end{equation}
are the Oseen tensor and the corresponding pressure Green's function.
As in
\cite{%
Bhattacharya-Blawzdziewicz-Wajnryb:2005,
Bhattacharya-Blawzdziewicz-Wajnryb:2005a%
}
the wall contributions to the Green's functions, $\wallGreenT$ and
$\wallGreenQ$, are represented in terms of lateral Fourier integrals
of simple matrix products.

It is convenient to express components of the tensor $\wallGreenT$ and
vector $\wallGreenQ$ in terms of the spherical basis of unit vectors
\cite{Edmonds:1960}
\begin{equation}
\label{spherical basis vectors}
\eSpherical_{-1}=\frac{1}{\sqrt{2}}(\ex-\im\ey),\quad\eSpherical_0=\ez,\quad
\eSpherical_{1}=-\frac{1}{\sqrt{2}}(\ex+\im\ey).
\end{equation}
Accordingly, we have
\begin{subequations}
\label{Expansion of Green's functions into spherical components}
\begin{equation}
\label{expansion of Green's function T into spherical components}
\wallGreenT(\br_1,\br_2)=\sum_{m=-1}^1\sum_{m'=-1}^1
    \wallGreenComponentT_{mm'}(\br_1,\br_2)\eSpherical_m\eSpherical_{m'}^*,
\end{equation}
\begin{equation}
\label{expansion of Green's function Q into spherical components}
\wallGreenQ(\br_1,\br_2)=\sum_{m'=-1}^1
   \wallGreenComponentQ_{m'}(\br_1,\br_2)\eSpherical_{m'}^*.
\end{equation}
\end{subequations}
The components $\wallGreenComponentT_{mm'}$ and
$\wallGreenComponentQ_{m'}$ can be evaluated from the following
2D Fourier integrals
\begin{widetext}
\begin{subequations}
\label{Fourier integrals for components of Green T and Q}
\begin{equation}
\label{Fourier integrals for components of Green T}
\wallGreenComponentT_{mm'}(\br_1,\br_2)
   =-\frac{\im^{m'-m}}{8\pi\eta}
     \int t_{mm'}(k;z_1,z_2)
     \e^{\im(m'-m)\psi}\e^{\im \bk\bcdot\brho_{12}}
     \frac{\diff\bk}{2\pi k},
\end{equation}
\begin{equation}
\label{Fourier integrals for components of Green Q}
\wallGreenComponentQ_{m'}(\br_1,\br_2)
   =-\frac{\im^{m'}}{4\pi}
     \int q_{m'}(k;z_1,z_2)
     \e^{\im m'\psi}\e^{\im \bk\bcdot\brho_{12}}
     \frac{\diff\bk}{2\pi k},
\end{equation}
\end{subequations}
\end{widetext}
where $\brho_{12}=\brho_1-\brho_2$, and $\bk=(k,\psi)$ represents the
wave vector $\bk$ in polar coordinates.  The integral kernels in Eqs.\
\refeq{Fourier integrals for components of Green T and Q} can be
expressed as products of several simple matrices,
\begin{subequations}
\label{integrands t and q}
\begin{equation}
\label{integrand t}
t_{mm'}(k;z_1,z_2)=
\STmatrix^\dagger(z_1,k,m)\bcdot\tildeZW(\bk)\bcdot\STmatrix(z_2,k,m'),
\end{equation}
\begin{equation}
\label{integrand q}
q_{m'}(k;z_1,z_2)
   =\Qmatrix^\dagger(z_1,k)\bcdot\tildeZW(\bk)\bcdot\STmatrix(z_2,k,m'),
\end{equation}
\end{subequations}
where the dagger denotes the transpose.

As explained in 
\cite{%
Bhattacharya-Blawzdziewicz-Wajnryb:2005,%
Bhattacharya-Blawzdziewicz-Wajnryb:2005a%
},
the  matrix 
\begin{equation}
\label{two wall Z matrix}
   \tildeZW(\bk)=
\left[
   \begin{array}{cc}
      \ZsingleWall^{-1}&\tildeCartesianDisplacement{++}(-kH)
\\\\
      \tildeCartesianDisplacement{--}(kH)&\ZsingleWall^{-1}
   \end{array}
\right]^{-1}
\end{equation}
describes the multiple reflections of Cartesian hydrodynamic basis
fields from the parallel walls.  The $3\times3$ component displacement matrices
\begin{equation}
\label{expression for tilde S}
\tildeCartesianDisplacement{++}(-kH)
   =[\tildeCartesianDisplacement{--}(kH)]^\dagger
=\left[
   \begin{array}{ccc}
      1&0&-2kH\\
      0&1&0\\
      0&0&1
   \end{array}
\right]\e^{-kH}
\end{equation}
describe the propagation of the flow fields between the walls, and the
matrices
\begin{equation}
\label{expression for Z wall}
\ZsingleWall
   =\left[
       \begin{array}{ccc}
          1&0&0\\
          0&1&0\\
          0&0&1
       \end{array}
    \right]
\end{equation}
represent scattering of the flow field from the walls.  The matrices
\begin{equation}
\label{matrix ST Lower Upper}
\STmatrix(k,z,m)=
[2(1-m)!(1+m)!]^{-1/2}\left[
\begin{array}{c}
\STmatrix^\low(k,z,m)\\
\STmatrix^\up(k,z,m)
\end{array}
\right]
\end{equation}
where
\begin{subequations}
\label{matrix ST}
\begin{equation}
\label{matrix ST lower}
\STmatrix^\low(k,z,m)=
(-1)^{m+1}\e^{-kz}
\left[
\begin{array}{c}
-2kz+2m^2-1\\
2m\\
1
\end{array}
\right],
\end{equation}
\begin{equation}
\label{matrix ST upper}
\STmatrix^\up(k,z,m)=
\e^{-k(H-z)}
\left[
\begin{array}{c}
1\\
2m\\
-2k(H-z)+2m^2-1
\end{array}
\right],
\end{equation}
\end{subequations}
describe the expansion of the Stokeslet into Cartesian basis fields
centered at the positions of the lower and upper wall.  Finally,
\begin{equation}
\label{matrix Q Lower Upper}
\Qmatrix(k,z)=
2^{1/2}k\left[
\begin{array}{c}
\Qmatrix^\low(k,z)\\
\Qmatrix^\up(k,z)
\end{array}
\right]
\end{equation}
where
\begin{subequations}
\label{matrix Q}
\begin{equation}
\label{matrix Q lower}
\Qmatrix^\low(k,z)=
\e^{-kz}
\left[
\begin{array}{c}
1\\
0\\
0
\end{array}
\right],
\end{equation}
\begin{equation}
\label{matrix Q upper}
\Qmatrix^\up(k,z)=
\e^{-k(H-z)}
\left[
\begin{array}{c}
0\\
0\\
1
\end{array}
\right],
\end{equation}
\end{subequations}
correspond to the pressure (at point $z$), associated with the Cartesian
basis fields centered at the lower or upper wall.  The three
components in the matrices \refeq{expression for tilde S},
\refeq{expression for Z wall}, \refeq{matrix ST}, and \refeq{matrix Q}
correspond to the pressure, vorticity, and potential basis solutions
of Stokes equations.

The 2D Fourier integrals \refeq{Fourier integrals for components
of Green T and Q} can be converted into the 1D Hankel transforms
by performing the angular integration with the help of the relation
\begin{equation}
\label{angular integral of Fourier mode}
(2\pi)^{-1}\int_0^{2\pi}
   \e^{\im m\psi}\e^{\im\bk\bcdot\brho_{12}}\diff\psi
   =\im^m\e^{\im m\varphi_{12}}\BesselJ_m(k\rho_{12}),
\end{equation}
where $\varphi_{12}$ is the polar angle of the vector $\brho_{12}$,
and $\BesselJ_m(x)$ is the Bessel function of the order $m$.  The
resulting expressions are
\begin{subequations}
\label{components of Green T and Q - Hankel integrals}
\begin{equation}
\label{components of Green T - Hankel integrals}
\wallGreenComponentT_{mm'}(\br_1,\br_2)
   =(-1)^{m'-m}\e^{\im(m'-m)\varphi_{12}}
     {\tilde t}_{mm'}(\br_1,\br_2),
\end{equation}
\begin{equation}
\label{components of Green Q - Hankel integrals}
\wallGreenComponentQ_{m'}(\br_1,\br_2)
   =(-1)^{m'}\e^{\im m'\varphi_{12}}
     {\tilde q}_{m'}(\br_1,\br_2),
\end{equation}
\end{subequations}
where
\begin{subequations}
\label{Hankel integrals for components of Green T and Q}
\begin{equation}
\label{Hankel integrals for components of Green T}
{\tilde t}_{mm'}(\br_1,\br_2)
   =-\frac{1}{8\pi\eta}
     \int_0^\infty t_{mm'}(k;z_1,z_2)\BesselJ_{m'-m}(k\rho_{12})\diff k,
\end{equation}
\begin{equation}
\label{Hankel integrals for components of Green Q}
{\tilde q}_{m'}(\br_1,\br_2)
   =-\frac{1}{4\pi}
     \int_0^\infty q_{m'}(k;z_1,z_2)\BesselJ_{m'}(k\rho_{12})\diff k.
\end{equation}
\end{subequations}

Substituting \refeq{components of Green T and Q - Hankel integrals}
into \refeq{Expansion of Green's functions into spherical components}
and using definitions \refeq{spherical basis vectors} of the spherical
basis vectors we get
\begin{subequations}
\label{Cartesian components of Green T and Q}
\begin{eqnarray}
\label{Cartesiancomponents of Green T}
\wallGreenT(\br_1,\br_2)
   &=&{\tilde t}_{11}(\br_1,\br_2)\lateralUnitTensor
     +{\tilde t}_{1\,\mbox{\scriptsize$-1$}}(\br_1,\br_2)
                  (\lateralUnitTensor-2\hat\brho_{12}\hat\brho_{12})
\nonumber\\
     &&+\sqrt{2}[{\tilde t}_{01}(\br_1,\br_2)\ez\hat\brho_{12}
     +{\tilde t}_{10}(\br_1,\br_2)\hat\brho_{12}\ez]\nonumber\\
     &&+{\tilde t}_{00}(\br_1,\br_2)\ez\ez,
\end{eqnarray}
\begin{equation}
\label{Cartesian components of Green Q}
\wallGreenQ_{m'}(\br_1,\br_2)=  
      \sqrt{2}{\tilde q}_1(\br_1,\br_2)\hat\brho_{12}
     +{\tilde q}_0(\br_1,\br_2)\ez.
\end{equation}
\end{subequations}

We note that relations \refeq{Expansion of Green's functions into
spherical components}--\refeq{Cartesian components of Green T and Q}
for the Green's functions $\GreenT$ and $\GreenQ$ are equivalent to
those derived by Jones \cite{Jones:2004}, but our expressions are more
transparent. In particular, the dependence of the integrands on the
variables $z_1$ and $z_2$ is clearly factored out, because $z$ appears
only in the matrices \refeq{matrix ST Lower Upper} and \refeq{matrix Q
Lower Upper}.  Moreover, our expressions can easily be adapted to
other boundary conditions at the walls (e.g., a fluid--fluid
interface) by simply replacing the reflection matrix $\ZsingleWall$.

\section{Transformation vectors 
$\frictionProjectionVector(A\mid lm\sigma)$ and
$\frictionProjectionVector(lm\sigma\mid A)$}
\label{Projection vectors X}

In this Appendix we give explicit expressions for the transformational
vectors $\frictionProjectionVector(A\mid lm\sigma)$ (where
$A=\transl,\rot,\pressure$), defined in Eqs.\ \refeq{multipolar
expressions for F T D} and \refeq{coefficients c for given particle
motion and parabolic flow}.  Formulas for translational and rotational
transformation vectors $\frictionProjectionVector(\transl\mid
lm\sigma)$, $\frictionProjectionVector(lm\sigma\mid\transl)$,
$\frictionProjectionVector(\rot\mid lm\sigma)$, and
$\frictionProjectionVector(lm\sigma\mid\rot)$ were derived in
\cite{Bhattacharya-Blawzdziewicz-Wajnryb:2005a}.  The transformation
vector $\frictionProjectionVector(lm\sigma\mid\pressure)$ (in a
slightly different notation) is given in
\cite{Bhattacharya-Blawzdziewicz-Wajnryb:2006a}.  The remaining
transformation vector $\frictionProjectionVector(\pressure\mid
lm\sigma)$ can be directly obtained from relations (50)--(52) in Ref.\
\cite{Bhattacharya-Blawzdziewicz-Wajnryb:2006}.

The transformation vectors $\frictionProjectionVector$ in Eqs.\
\refeq{multipolar expressions for F T D} are given by the 
expressions
\begin{subequations}
\label{force transformation vectors}
\begin{equation}
\label{force transformation vector t}
\frictionProjectionVector(\transl\mid lm\sigma)
   =(\smallfrac{4}{3}\pi)^{1/2}\delta_{l1}\delta_{\sigma0}\eSpherical_m
\end{equation}
\begin{equation}
\label{force transformation vector r}
\frictionProjectionVector(\rot\mid lm\sigma)
   =-2\im(\smallfrac{4}{3}\pi)^{1/2}\delta_{l1}\delta_{\sigma1}\eSpherical_m
\end{equation}
\begin{equation}
\label{force transformation vector p}
\frictionProjectionVector(\pressure\mid lm\sigma)
   =2^{-1/2}mC(Z;lm\sigma)\eSpherical_m,
\end{equation}
\end{subequations}
where $\eSpherical_m$ denote the basis vectors \refeq{spherical basis
vectors}, and $m=-1,0,1$.  For other values of $m$, the transformation
vectors vanish.  It can be verified that the reciprocal transformation
vectors $\frictionProjectionVector(lm\sigma\mid A)$ are related to
\refeq{force transformation vectors} via the symmetry \refeq{symmetry
of vectors X}.

The transformation vector
$\frictionProjectionVector(\pressure\mid lm\sigma)$ is nonzero only
for $m=\pm1$ and
\begin{equation}
\label{condition for nonzero elements of C matrix}
l+\sigma\le3.
\end{equation} 
The coefficient $C$ in relation \refeq{force transformation vector p}
can be expressed in the form
\cite{Bhattacharya-Blawzdziewicz-Wajnryb:2006}
\begin{equation}
\label{nonzero elements of C matrix}
\TransformationElementSAs(Z;l\,\,\mbox{$\pm1$}\,\,\sigma)=
   \matrixElementBPM{l-1\,\,\sigma}(1;Z),
\end{equation}
where $\matrixElementBPM{\lambda\,\sigma}(1;Z)$ denote the elements
of the $3\times3$ matrix 
\begin{widetext}
\begin{equation}
\label{expression for matrix B}
\left\{
   \matrixElementBPM{\lambda\,\sigma}(1;Z)
\right\}_{\lambda,\sigma=0,1,2}
=\mp \left(\frac{2\pi}{3}\right)^{1/2}
   \left[
       \begin{array}{ccc}
          -Z(H-Z)&\mp(H-2Z)&2\\
&&\\
          \displaystyle\frac{-(H-2Z)}{2\sqrt{5}}
                    &\pm\displaystyle\frac{1}{\sqrt{5}}
                   &0\\
&&\\
          \displaystyle\frac{2}{15\sqrt{3}}
                   &0&0
       \end{array}
    \right].
\end{equation}
\end{widetext}
The range $\lambda=0,1,2$ of the index $\lambda=l-1$ in equation
\refeq{expression for matrix B} results from the conditions $l\ge1$
and \refeq{condition for nonzero elements of C matrix}.  

\section{Far-field contributions to periodic Green's functions}
\label{Periodic formulation for two-dimensional Laplace's equation}

In this Appendix we list explicit formulas for the far-field contributions to
periodic Green's functions for Stokes flow between two parallel walls.  In
Sec.\ \ref{Multipolar solutions of 2D Laplace equation} we first consider the
2D scalar problem.  We provide an Ewald-sum representation for
the Wigner function $\Wigner$ and list the corresponding expressions for the
periodic multipolar solutions of Laplace equation in 2D.  In Sec.\
\ref{Asymptotic Green's functions for Stokes flow} we give explicit formulas
for the periodic Green's function \refeq{periodic asymptotic Green's
functions} and for their spherical matrix elements \refeq{matrix elements}.

\subsection{Multipolar periodic solutions of 2D Laplace equation}
\label{Multipolar solutions of 2D Laplace equation}

The relations presented in this sections are based on the results of
Cichocki and Felderhof \cite{Cichocki-Felderhof:1989a} who have
derived Ewald-type expressions for the 2D periodic multipolar
potentials.

\subsubsection{Wigner function}

As shown in \cite{Cichocki-Felderhof:1989a}, Wigner function can be
represented by the following formula
\begin{widetext}
\begin{equation}
\label{Ewald sum for Wigner function}
\Wigner(\brho)
   =\frac{1}{2}\sum_\bn E_1\left(
      \frac{\pi|\brho-\brho_\bn|^{2}}{\sigma^2}
   \right)
   +\frac{1}{2\pi L_x L_y}\sum_{\bn\not=0}
      \frac{1}{k_\bn^2}
      \exp(-\pi\sigma^{2}k_\bn^2+2\pi\im{\bf r}\cdot{\bf k}_\bn)
   -\frac{\sigma^{2}}{2L_x L_y}+C_w.
\end {equation}
\end{widetext}
Here 
\begin{subequations}
\label{lattice vectors}
\begin{equation}
\label{direct lattice vectors}
\brho_\bn=n_x L_x \ex +n_y L_y \ey,
\end{equation}
\begin{equation}
\label{reciprocal lattice vectors}
\bk_\bn=\frac{n_x}{L_x}\ex +\frac{n_y}{L_y}\ey
\end{equation}
\end{subequations}
are the direct and reciprocal lattice vectors, 
\begin{equation}
\label{exponential function}
E_1(x)=\int_1^{\infty}\frac{\e^{-xt}}{t}\diff t
\end {equation}
is the exponential function, and $\sigma$ is the splitting parameter
that controls the convergence of the direct and reciprocal sums.  The
gauge constant $C_w=1.3105329259$ \cite{Cichocki-Felderhof:1989a} is
used to set the limit
\begin{equation}
\label{limit of Wigner function}
\lim_{\brho\to\brho_\bn}[\Wigner(\rho)-\Coulomb(\brho-\brho_\bn)]=0.
\end{equation}

\subsubsection{Multipolar solutions}

The periodic multipolar solutions $\periodicScalarBasisM{m}$ of the 2D
Laplace equation are defined in terms of the non-periodic multipolar
basis fields
\begin{subequations}
\label{Scalar basis fields}
\begin{equation}
\label{Scalar basis minus}
\ScalarBasisM{0}(\lateralVector)=-\ln\lateralDistance,\qquad
\ScalarBasisM{m}(\lateralVector)=\frac{1}{2|m|}
   \lateralDistance^{-|m|}\e^{\im m\phi}, \quad m\not=0,
\end{equation}
\begin{equation}
\label{Scalar basis plus}
\ScalarBasisP{m}(\lateralVector)=
   \lateralDistance^{|m|}\e^{\im m\phi}.
\end{equation}
\end{subequations}
By definition, for $m=0$ we simply have
\begin{equation}
\label{multipolar field 0}
\periodicScalarBasisM{0}(\brho)\equiv\Wigner(\brho).
\end{equation}
For nonzero values of $m$, the multipolar solutions are given by the
expression
\begin{widetext}
\begin{equation}
\label{periodic phi m}
\periodicScalarBasisM{m}(\brho)
   =\frac{1}{\Gamma(|m|)}
    \sum_\bn
      \Gamma(|m|,\pi\sigma^{-2}|\brho-\brho_\bn|^{2})
      \ScalarBasisM{m}(\brho-\brho_\bn)
  +\frac
        {\pi^{|m|-1}\,\im^{|m|}}
        {2|m|!L_x L_y}
  \sum_{\bn\not=0}k_\bn^{-2}
\ScalarBasisP{m}(\bk_\bn)
\exp(-\pi\sigma^{2}k_\bn^{2}+2\pi\im\brho\bcdot\bk_\bn),
\end {equation}
\end{widetext}
where $\Gamma(j,x)$ is the incomplete Gamma function.

Near the singularities at the lattice points $\brho=\brho_\bn$ the
periodic functions \refeq{periodic phi m} behave as
\begin{equation}
\label{asymptotic behavior of periodic multipoles}
\periodicScalarBasisM{m}(\brho)
   \simeq\ScalarBasisM{m}(\brho-\brho_\bn).
\end{equation}
For $|m|>0$ the gauge constants are determined by the condition that
the fields $\periodicScalarBasisM{m}(\brho)$ can be expressed as
combinations of derivatives of the Wigner function $\Wigner(\brho)$.

\subsubsection{Displacement theorems}

The multipolar fields \refeq{periodic phi m} can be used to construct
the displacement formula for the Wigner potential
\cite{Cichocki-Felderhof:1989a}
\begin{equation}
\label{displacement formula for Wigner's potential}
\Wigner(\brho+\bar\brho)
  =\frac{\pi}{2L_x L_y}\bar\rho^2
  +\sum_{m_1=-\infty}^{\infty}\periodicScalarBasisM{m_1}(\brho)
                            \ScalarBasisPcon{m_1}(\bar\brho),
\end{equation}
where the first term on the right-hand side corresponds to the
background term in the periodic Poisson equation \refeq{periodic
Poisson equation}.  Setting $\bar\brho=\brho'-\brho_{12}$ and using
relations
\begin{widetext}
\begin{equation}
\label{decomposition of phi+}
\ScalarBasisPcon{m_1}(\brho-\brho')
   =\sum_{\stackrel{\scriptstyle m,m'}{m+m'=m_1}}
    (-1)^{m}\theta(mm')
    \frac{(|m|+|m'|)!}{|m|!\,|m'|!}
    \ScalarBasisP{-m}(\brho)\ScalarBasisPcon{m'}(\brho'),
\end{equation}
\end{widetext}
[where $\theta(x)$ is the Heaviside step function] and
\begin{equation}
\label{product of rho}
2\brho\bcdot\brho'
   =\ScalarBasisP{1}(\brho)\ScalarBasisPcon{1}(\brho')
          +\ScalarBasisP{-1}(\brho)\ScalarBasisPcon{-1}(\brho')
\end{equation}
we find a symmetric displacement relation 
\begin{widetext}
\begin{equation}
\label{displacement formula for Wigner's potential expanded}
\Wigner(\brho+\brho_{12}-\brho')
  =\frac{\pi}{2L_x L_y}(\rho^2+\rho^{\prime\,2})
  +\sum_{m=-\infty}^{\infty}\sum_{m'=-\infty}^{\infty}
         \ScalarBasisP{m}(\brho)
         \periodicScalarDisplacementElementsPM(\brho_{12};m\mid m')
                            \ScalarBasisPcon{m'}(\brho'),
\end{equation}
where 
\begin{equation}
\label{2D electrostatic displacement elements}
\periodicScalarDisplacementElementsPM(\brho_{12};m\mid m')
  =-\frac{\pi}{2L_x L_y}(\delta_{1m}\delta_{1m'}+\delta_{-1m}\delta_{-1m'})
  +\theta(-mm')(-1)^{m'}
     \frac{(|m|+|m'|)!}{|m|!|m'|!}
     \periodicScalarBasisM{m'-m}(\brho_{12})
\end{equation}
\end{widetext}
is the displacement matrix for periodic multipolar scalar fields.
Integrating the above expressions with the multipolar source
distribution of order $m'$, centered at $\rho'=0$ yields the
displacement theorem for the multipolar periodic solutions
\refeq{periodic phi m},
\begin{equation}
\label{2D electrostatic displacement theorem}
\periodicScalarBasisM{m'}(\brho+\brho_1)
   =\delta_{m0}\frac{\pi}{2L_x L_y}\rho_1^2+
    \sum_{m=-\infty}^\infty \ScalarBasisP{m}(\brho_1)
    \periodicScalarDisplacementElementsPM(\brho;m\mid m'),
\end{equation}
where the convergence condition $\rho_1<\rho$ is assumed.

\subsection{Asymptotic Green's functions for Stokes flow}
\label{Asymptotic Green's functions for Stokes flow} 

\subsubsection{Green's functions {\rm$\periodicGreenTasymptotic$} and 
{\rm $\periodicGreenQasymptotic$}}

The relations given in the previous section can be used to derive explicit
expressions for the asymptotic Green's functions \refeq{periodic asymptotic
Green's functions}.  Taking the gradients of relation \refeq{displacement
formula for Wigner's potential expanded} with respect to variables $\brho$
and $\brho'$, and evaluating the results at $\brho=\brho'=0$ we find
\begin{widetext}
\begin{subequations}
\label{periodic Green's functions T and Q in terms of scalar basis}
\begin{equation}
\label{periodic Green's function T  in terms of scalar basis}
\periodicGreenTasymptotic(\br_1,\br_2)
  =-\frac{3}{\pi\eta H^3}z_1(H-z_1)z_2(H-z_2)
    \sum_{m=-1,1}\sum_{m'=-1,1}mm'
    \periodicScalarDisplacementElementsPM(\brho_{12};m\mid m')
                           \eSpherical_m\eSpherical_{m'}^*\,,
\end{equation}
\begin{equation}
\label{periodic Green's function Q  in terms of scalar basis}
\periodicGreenQasymptotic(\br_1,\br_2)
  =-\frac{3\sqrt{2}}{\pi H^3}z_2(H-z_2)
    \sum_{m=-1,1}m'
    \periodicScalarDisplacementElementsPM(\brho_{12};0\mid m')
                           \eSpherical_{m'}^*\,.
\end{equation}
\end{subequations}
\end{widetext}
The sums in \refeq{periodic Green's functions T and Q in terms of
scalar basis} can be evaluated explicitly using expressions
\refeq{periodic phi m}, \refeq{2D electrostatic displacement
elements}, and \refeq{non-periodic asymptotic Green's functions},
\begin{widetext}
\begin{subequations}
\label{Elek's formulas for periodic T and Q}
\begin{eqnarray}
\label{Elek's formulas for periodic T}
\periodicGreenTasymptotic(\br_1,\br_2)
   =\frac{3}{2\eta H^3L_x L_y}z_1(H-z_1)z_2(H-z_2)\lateralUnitTensor
   +\sum_\bn\Gamma(2,\pi\sigma^{-2}|\brho_{12}-\brho_\bn|^2)
            \GreenTasymptotic(\brho_{12}-\brho_\bn;z_1,z_2)
&&\nonumber\\
   -\frac{4\pi}{L_x L_y}\sum_{\bn\not=0}
            k_\bn^2\GreenTasymptotic(\bk_\bn;z_1,z_2)
            \exp(-\pi\sigma^{2}k_\bn^2+2\pi\im\brho_{12}\bcdot\bk_\bn),
\rule{118pt}{0pt}
&&
\end{eqnarray}
\begin{equation}
\label{Elek's formulas for periodic Q}
\periodicGreenQasymptotic(\br_1,\br_2)
   =\sum_\bn\Gamma(1,\pi\sigma^{-2}|\brho_{12}-\brho_\bn|^2)
            \GreenQasymptotic(\brho_{12}-\brho_\bn;z_2)
   +\frac{2\im}{L_x L_y}\sum_{\bn\not=0}
            \GreenQasymptotic(\bk_\bn;z_2)
            \exp(-\pi\sigma^{2}k_\bn^2+2\pi\im\brho_{12}\bcdot\bk_\bn),
\end{equation}
\end{subequations}
\end{widetext}
where we have introduced notation
\begin{subequations}
\begin{equation}
\label{notation for Green T}
\GreenTasymptotic(\brho_{12};z_1,z_2) =
\GreenTasymptotic(\br_1,\br_2),
\end{equation}
\begin{equation}
\label{notation for Green Q}
\GreenQasymptotic(\brho_{12};z_1,z_2) =
\GreenQasymptotic(\br_1,\br_2).
\end{equation}
\end{subequations}
The quickly convergent Ewald sums \refeq{Elek's formulas for periodic
T and Q} can be used for efficient evaluation of the periodic Green's
functions $\periodicGreenTasymptotic$ and $\periodicGreenQasymptotic$
in Stokesian-dynamics and boundary-integral applications.

\subsubsection{Matrix elements}

The projections
\begin{equation}
\label{matrix elements - periodic, asymptotic}
\GreenWallTotElement_\asymptotic^\periodic
                     (lm\sigma;\br_1\mid l'm'\sigma';\br_2)
        =\langle\reciprocalSphericalBasisP{lm\sigma}(\br_1)
\mid
   \periodicGreenTasymptotic
\mid
   \reciprocalSphericalBasisP{l'm'\sigma'}(\br_2)\rangle,
\end{equation}
of the periodic Hele--Shaw Green's function
$\periodicGreenTasymptotic$ onto the 3D spherical basis can be
obtained using relation \refeq{displacement formula for Wigner's
potential expanded} and applying the method described in
\cite{Bhattacharya:2005}.  The results can be written in the form
analogous to Eq.\ (46) in Ref.\
\cite{Bhattacharya-Blawzdziewicz-Wajnryb:2006},
\begin{widetext}
\begin{equation}
\label{expressions for matrix elements - periodic, asymptotic}
\GreenWallTotElement_\asymptotic^\periodic
                     (lm\sigma;\br_1\mid l'm'\sigma';\br_2)
=-\frac{6}{\pi\eta H^3}
       C(Z_1;lm\sigma)
       \periodicScalarDisplacementElementsPM(\brho_{12};m\mid m')
       C(Z_2;l'm'\sigma').
\end{equation}
\end{widetext}


\begin{thebibliography}{10}
\bibitem{scalar_coefficients}
For a single spherical particle, the system has a cylindrical symmetry, and the
  two-dimensional tensors $\mobilityMatrix^{AB}$ can be replaced with scalar
  coefficients. However, tensorial notation is needed for multiparticle
  configurations and for systems of non-spherical particles.
\bibitem{convergence_note}
Strictly speaking, for $m=\pm1$ the lattice sums \refeq{direct sums for
  asymptotic periodic Green's functions} are not absolutely convergent.
  However, Eqs.\ \refeq{periodic asymptotic Green's functions} and
  \refeq{periodic Poisson equation} yield a unique definition of the functions
  $\periodicGreenTasymptotic$ and $\periodicGreenQasymptotic$.
\end{thebibliography}

\end{document}